\documentclass[12pt,preprint]{aastex}

\usepackage{graphicx}
\usepackage{txfonts}
\newcommand{\va}{v_{\mathrm{A}}}
\newcommand{\vaf}{v_{\mathrm{Ap}}}
\newcommand{\vac}{v_{\mathrm{Ac}}}
\newcommand{\vae}{v_{\mathrm{Ae}}}

\newcommand{\vazero}{v_{\mathrm{A}_0}}

\newcommand{\pd}{\partial}
\newcommand{\td}{\tau_{\mathrm{D}}}
\newcommand{\tdp}{\tau_{\mathrm{D}} / P}

\newcommand{\mutilde}{\tilde \mu}
\newcommand{\mut}{\tilde \mu}
\newcommand{\mutildef}{\tilde{\mu}_{\rm p}}
\newcommand{\mutildec}{\tilde{\mu}_{\rm c}}

\newcommand{\etacf}{\eta_{\rm Cp}}
\newcommand{\etacc}{\eta_{\rm Cc}}
\newcommand{\etace}{\eta_{\rm Ce}}

\newcommand{\etactf}{\tilde{\eta}_{\rm Cp}}
\newcommand{\etactc}{\tilde{\eta}_{\rm Cc}}

\newcommand{\etaczero}{\eta_{{\rm C}_0}}

\begin{document}

	\title{SEISMOLOGY OF STANDING KINK OSCILLATIONS OF SOLAR PROMINENCE FINE STRUCTURES}

	\shorttitle{STANDING KINK OSCILLATIONS OF PROMINENCE FINE STRUCTURES}

   \author{R. Soler, I. Arregui, R. Oliver, and J. L. Ballester}
   \affil{Departament de F\'isica, Universitat de les Illes Balears,
              E-07122, Palma de Mallorca, Spain}
              \email{roberto.soler@uib.es}

  \begin{abstract}

We investigate standing kink magnetohydrodynamic (MHD) oscillations in a prominence fine structure modeled as a straight and cylindrical magnetic tube only partially filled with the prominence material, and with its ends fixed at two rigid walls representing the solar photosphere. The prominence plasma is partially ionized and a transverse inhomogeneous transitional layer is included between the prominence thread and the coronal medium. Thus, ion-neutral collisions and resonant absorption are the considered damping mechanisms. Approximate analytical expressions of the period, the damping time, and their ratio are derived for the fundamental mode in the thin tube and thin boundary approximations. We find that the dominant damping mechanism is resonant absorption, which provides damping ratios in agreement with the observations, whereas ion-neutral collisions are irrelevant for the damping. The values of the damping ratio are independent of both the prominence thread length and its position within the magnetic tube, and coincide with the values for a tube fully filled with the prominence plasma. The implications of our results in the context of the MHD seismology technique are discussed, pointing out that the reported short-period (2 -- 10~min) and short-wavelength (700 -- 8,000~km) thread oscillations may not be consistent with a standing mode interpretation and could be related to propagating waves. Finally, we show that the inversion of some prominence physical parameters, e.g., Alfv\'en speed, magnetic field strength, transverse inhomogeneity length-scale, etc., is possible using observationally determined values of the period and damping time of the oscillations along with the analytical approximations of these quantities.

  \end{abstract}



\section{INTRODUCTION}

Oscillations and propagating waves are commonly reported in observations of solar prominences and filaments \citep[see recent reviews by, e.g.,][]{ballester,engvold,mackay}. In high-resolution observations, transverse oscillations of prominence fine structures are frequently detected. These fine structures, here called threads, appear as a myriad of long ($5'' - 20''$) and thin ($0''.2 - 0''.6$) dark ribbons in H$\alpha$ images of filaments on the solar disk \citep[e.g.,][]{lin07,lin08,lin09}, as well as in observations of prominences in the solar limb from the Solar Optical Telescope (SOT) aboard the Hinode satellite \citep[e.g.,][]{okamoto,berger,chae,ning}. From the theoretical point of view, prominence fine structures have been modeled as magnetic flux tubes anchored in the solar photosphere \citep[e.g.,][]{ballesterpriest,rempel}, which are piled up to form the prominence body. In this interpretation, only part of the flux tubes would be filled with the cool ($\sim 10^4$~K) filament material, which would correspond to the observed threads, while the rest of the magnetic tube, i.e., the so-called evacuated zone, would be occupied by hot coronal plasma. 

 Common features of the transverse oscillations of prominence fine structures detected in Doppler signals and H$\alpha$ sequences are that the reported periods are usually in a narrow range between 2 and 10 minutes, that the velocity amplitudes are smaller than $\sim 3$~km~s$^{ -1}$, and that the oscillations seem to be damped after a few periods. Typically, the number of oscillatory periods observed before the oscillations disappear is less than 10 \citep[see, e.g.,][]{molowny99,terradasobs,lin04,ning}. Theoretically, the oscillations have been interpreted in terms of kink magnetohydrodynamic (MHD) modes supported by the fine structure, modeled as a magnetic slab  \citep{joarder97,diaz2001,diaz2003} or a cylindrical tube \citep{diaz2002,diazperiods,dymovaruderman,hinode} partially filled with the prominence plasma, whereas several damping mechanisms have been proposed to explain the quick attenuation \citep[see, e.g.,][]{oliver,arreguiballester,solerphd}. 

By neglecting the variation of the plasma parameters along the fine structure and adopting a prominence thread model composed of a homogeneous magnetic flux tube with prominence conditions embedded in a coronal environment, \citet{solercylnoad} studied the temporal damping of propagating kink MHD waves due to nonadiabatic effects (radiative losses, thermal conduction, and plasma heating), while \citet{solerneutrals} investigated the attenuation in the same configuration but considering ion-neutral collisions as the damping mechanism. These authors concluded that neither nonadiabatic effects nor ion-neutral collisions can produce kink mode damping times compatible with those observed. On the other hand, \citet{arregui} considered a similar model but neglected gas pressure (i.e., the $\beta = 0$ approximation, with $\beta$ the ratio of the gas pressure to the magnetic pressure) and took into account the presence of a transversely inhomogeneous transitional layer between the thread and the coronal plasma. In such a configuration, the kink mode is resonantly coupled to Alfv\'en continuum modes, and so the kink mode is damped by the process of resonant absorption. \citet{arregui} numerically obtained values of the damping time that are consistent with those reported in the observations. Resonant absorption has been previously proposed as an explanation for both the temporal damping of coronal loop transverse oscillations \citep[e.g.,][]{rudermanroberts02,goossens02} and the spatial damping of propagating kink waves \citep[e.g.,][]{pascoe,terradasspatial}.  Subsequently, \citet{solerslow} performed a more in-depth analytical and numerical investigation of the damping by resonant absorption in prominence threads by including gas pressure, and obtained similar results to those of \citet{arregui}. More recently, \citet{solerRAPI} studied the combined effect of resonant absorption and the prominence plasma partial ionization on the kink mode damping. For realistic values of the wavelength, \citet{solerRAPI} concluded that partial ionization does not affect the process of resonant absorption, and so the obtained values of the damping time are the same as in a fully ionized prominence thread \citep{arregui,solerslow}. Ion-neutral collisions may become more relevant than resonant absorption for the temporal damping of propagating kink modes when short wavelengths of the order of 10$^3$~km and smaller values are considered, while the range of typically observed wavelengths in prominences is between $5\times 10^3$ -- $10^5$~km \citep{oliverballester}.

Whether the reported observations of transversely oscillating filament and prominence threads are related to propagating waves or standing oscillations is a subject of debate. For example, \citet{lin09} explained their observations in terms of propagating kink waves, whereas \citet{hinode} proposed standing oscillations as an explanation of the observations by \citet{okamoto}. It is likely that propagating waves may be generated by localized disturbances in, e.g., the footpoints of the magnetic tube,  while standing oscillations may be related to more global perturbations of the whole magnetic structure. Regarding the damping, all the works cited above studied the temporal damping of propagating kink waves, hence the wavelength (or the wavenumber) is a free parameter in their case, but the problem of the damping of standing oscillations has not been addressed yet in the context of prominence fine structures. Moreover, the effect on the damping of the longitudinal variation of the plasma parameters along the fine structure was not addressed in these previous works. Therefore, the aim of the present investigation is to broach the problem of the damping of standing kink MHD oscillations of longitudinally nonuniform prominence fine structures.

 The model configuration adopted here is similar to that considered by \citet{diaz2002,diazperiods}, \citet{dymovaruderman}, and \citet{hinode}, namely a straight and cylindrical magnetic flux tube only partially filled with prominence plasma. The rest of the magnetic tube, as well as the external medium, has typical coronal properties. The $\beta = 0$ approximation is adopted for the sake of simplicity since we restrict ourselves to kink modes, which are correctly described by this approximation. Standing oscillations are studied by imposing the line-tying condition at the ends of the cylinder. As in \citet{solerRAPI}, we assume that the prominence plasma is partially ionized and include a transversely inhomogeneous transitional layer between the dense prominence thread and the external corona. Hence, the mechanisms of ion-neutral collisions and resonant absorption are considered as damping mechanisms. We follow the method introduced by \citet{dymovaruderman} based on the thin tube limit, and derive a dispersion relation for damped kink MHD oscillations in the thin boundary approximation. Analytical expressions of the period, the damping time, and the ratio of the damping time to the period are obtained, while a general parametric study is performed by numerically solving the full dispersion relation. In a subsequent work, 	\citet{arregui2d} investigate the damping of kink oscillations beyond the thin tube and thin boundary approximations by numerically solving the full resistive eigenvalue problem in two-dimensional, nonuniform threads.

This paper is organized as follows. Section~\ref{sec:math} includes a description of the model configuration and the mathematical method. The dispersion relation of standing kink MHD oscillations is obtained in Section~\ref{sec:disper}, which also contains analytical approximations. The results of solving the dispersion relation are given in Section~\ref{sec:results}, while their implications for prominence seismology are discussed in Section~\ref{sec:seismology}. Finally, Section~\ref{sec:conclusion} contains our main conclusions.

\section{MODEL AND METHOD}
\label{sec:math}

\subsection{Equilibrium Configuration}
\label{sec:equi}

\begin{figure*}[!htp]
\centering
 \epsscale{0.9}
 \plotone{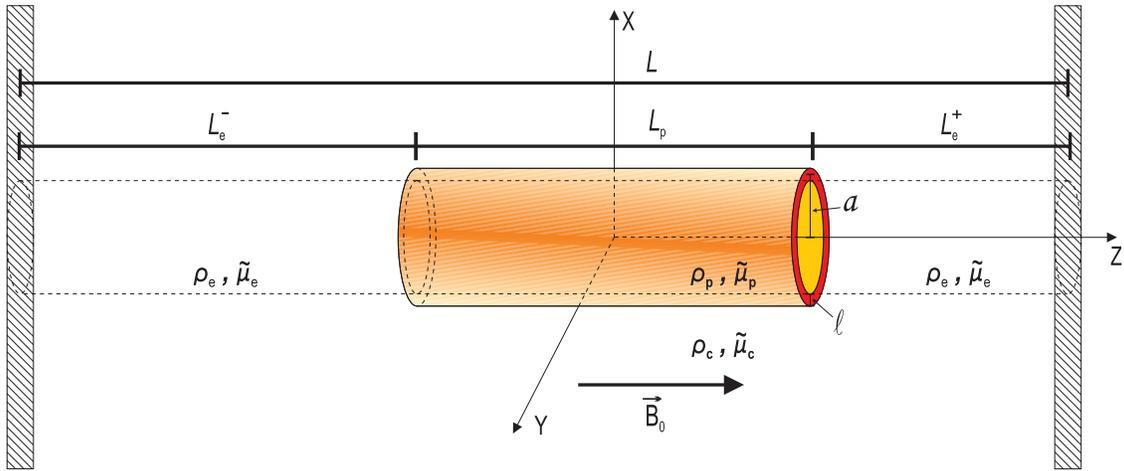}
\caption{Sketch of the model configuration adopted in this work. \label{fig:model}}
\end{figure*}

The model considered here is schematically plotted in Figure~\ref{fig:model}. We consider a straight and cylindrical magnetic tube of length $L$ and radius $a$, whose ends are fixed by two rigid walls representing the solar photosphere. The magnetic tube is only partially filled with the cool and dense prominence material, and is composed of a  dense region of length $L_{\rm p}$ with prominence conditions and representing the prominence thread, surrounded by two much less dense zones corresponding to the evacuated part of the tube. According to the observed typical values of thread widths and lengths from the high-resolution observations \citep[e.g.,][]{lin04,lin08}, the ranges of realistic values of $a$ and $L_{\rm p}$ are 50~km~$\lesssim a \lesssim$~300~km and 3,000~km~$\lesssim L_{\rm p} \lesssim$~28,000~km. On the other hand, the total tube length, $L$, cannot be measured from the observations, but one can relate $L$ to the typical spatial scale in prominences and filaments, i.e., $L \sim 10^5$~km.

For simplicity,  both the prominence and the evacuated (i.e., coronal) part are taken homogeneous, with densities $\rho_{\rm p} $ and $\rho_{\rm e}$, respectively. The external, coronal medium has density $\rho_{\rm c}$, which is also homogeneous. Subscripts p, e, and c denote the prominence part, the evacuated region, and the corona, respectively. In general, the subscript 0 denotes equilibrium quantities without referring to a particular region. The density contrast of the prominence part with respect to the coronal plasma is a large parameter, with $\rho_{\rm p} / \rho_{\rm c} = 200$ a value usually considered. We assume that the evacuated part of the tube has the same density as the corona. Hence, for a typical prominence density of $\rho_{\rm p} = 5 \times 10^{-11}$~kg~m$^{-3}$, the coronal and evacuated densities are $\rho_{\rm c} = \rho_{\rm e} = 2.5 \times 10^{-13}$~kg~m$^{-3}$. In the prominence region, we include a transversely inhomogeneous transitional layer of thickness $l$, that continuously connects the internal prominence region to the external corona. A sinusoidal variation of the density is considered in the transitional layer \citep{rudermanroberts02}. The limits $l/a = 0$ and $l/a=2$ correspond to a thread without transverse transitional layer and a fully inhomogeneous thread in the radial direction, respectively. The plasma in the prominence region is assumed partially ionized with an arbitrary ionization degree $\mutildef$. Both the evacuated part and the corona are taken fully ionized, hence $\mutilde_{\rm e} = \mutildec = 0.5$.

 We use cylindrical coordinates, namely $r$, $\varphi$, and $z$ for the radial, azimuthal, and longitudinal coordinates. The magnetic field is taken homogeneous and orientated along the $z$-direction, namely ${\bf B}_0=B_0\hat{e}_z$, with $B_0=5$~G everywhere. The $z$-direction also coincides with the axis of the cylinder. For $l/a=0$ and full ionization, the model is equivalent to that assumed by \citet{diaz2002} and \citet{dymovaruderman}. In these two works the prominence thread is located in the center of the cylinder. Here, we allow the thread to be displaced from the center of the tube. The length of the evacuated region on the left-hand side of the thread is $L_{\rm e}^-$, whereas the length of the right-hand side evacuated region is $L_{\rm e}^+$.  When $L$ and $L_{\rm p}$ are fixed, we can express $L_{\rm e}^+ = L - L_{\rm e}^- - L_{\rm p}$, hence it is enough to select a value for $L_{\rm e}^-$ in order to set the length of both evacuated parts. The allowed values of $L_{\rm e}^-$ are in the range $0 \leq L_{\rm e}^- \leq L - L_{\rm p}$. For $L_{\rm e}^- = 0$, the thread is totally displaced to the left-hand side end of the flux tube, while the contrary occurs for $L_{\rm e}^- = L - L_{\rm p}$. For $L_{\rm e}^- = L_{\rm e}^+ = \frac{1}{2} \left( L - L_{\rm p} \right)$ the prominence thread is located at the center of the tube, i.e., the configuration studied by \citet{diaz2002} and \citet{dymovaruderman}. For simplicity, we fix the origin of coordinates at the center of the prominence region, so that the interfaces between the prominence plasma and the evacuated zones are located at  $z = \pm L_{\rm p}/2$. The photospheric walls are therefore located at $z=z^-_{\rm wall} = - L_{\rm p}/2  - L_{\rm e}^-$ and  $z=z^+_{\rm wall} = L_{\rm p}/2 +  L_{\rm e}^+$, with $z^-_{\rm wall}$ and $z^+_{\rm wall}$ the position of the left and right walls, respectively. Thus, for $L_{\rm p}=L$, i.e, a tube fully filled with prominence material, the model reduces to the configuration studied by \citet{solerRAPI}.

\subsection{Basic Equations}

The governing MHD equations for a partially ionized plasma are derived in \citet{forteza07} after the single-fluid treatment of \citet{brag} \citep[see also details in][]{pinto,solerneutrals,solerphd}. Electrons, ions (i.e., protons), and neutral hydrogen are the species taken into account. We assume small-amplitude perturbations over the equilibrium state and the basic equations are linearized \citep[][Equations~(4)--(7)]{solerneutrals}. In a partially ionized plasma, the induction equation contains diffusion terms related to the collisions between the different species \citep[see][]{pinto}. For example, Ohm's diffusion is governed by electron-ion collisions, whereas Cowling's diffusion is dominated by ion-neutral collisions. Here, we neglect Ohm's diffusion from the induction equation \citep[][Equation~(7)]{solerneutrals} because its role in a partially ionized prominence plasma is much less important than that of Cowling's diffusion. In addition, since we adopt the $\beta = 0$ approximation, gas pressure effects are neglected.

Next, we follow \citet{forteza08} and \citet{solerneutrals} and take  a time dependence of the form $\exp \left( - i \omega t \right)$, with $\omega$ the oscillatory frequency, so that we can include the effect of Cowling's diffusion in the definition of a modified Alfv\'en speed squared as $\Gamma_{\rm A_0}^2 = \vazero^2 - i \omega \etaczero$, where $\vazero^2 =  B_0^2 / \mu  \rho_0$ is  the Alfv\'en speed squared and $\etaczero$ is the Cowling's diffusion coefficient, with $\mu = 4 \pi  \times 10^{-7}$~N~A$^{-2}$. The expression of Cowling's diffusivity, $\etaczero$, in terms of the equilibrium properties is given in, e.g., \citet{solerneutrals} and \citet{solerphd}. Thus, the relevant equations for our investigation are
\begin{equation}
 \rho_0  \frac{\partial {\bf v}_1}{\partial t} = \frac{1}{\mu}\left( \nabla \times {\bf B}_1 \right) \times {\bf B}_0, \label{eq:motionbeta0app} 
\end{equation}
\begin{equation}
  \frac{\partial {\bf B}_1}{\partial t} = \frac{\Gamma_{\rm A_0}^2}{\vazero^2} \nabla \times \left( {\bf v}_1 \times {\bf B}_0\right), \label{eq:inductionapp}
\end{equation}
where ${\bf v}_1 = \left( v_r, v_\varphi, v_z \right)$ and ${\bf B}_1 = \left( B_r, B_\varphi, B_z \right)$ are the velocity and the magnetic field perturbations. Note that $v_z = 0$ in the $\beta = 0$ approximation, and $\Gamma_{\rm A_0}/ \vazero = 1$ when Cowling's diffusion is absent.

Equations~(\ref{eq:motionbeta0app}) and (\ref{eq:inductionapp}) can be combined to arrive at the following equation for the total pressure perturbation, $p_{{\rm T}} = B_0 B_z / \mu$, namely
\begin{equation}
 \frac{\pd^2 p_{{\rm T}}}{\pd t^2} - \Gamma_{\rm A_0}^2 \nabla^2 p_{{\rm T}} = 0, \label{eq:pt11}
\end{equation}
along with an equation relating the total pressure and radial velocity perturbations as
\begin{equation}
  \frac{\pd^2 v_r}{\pd t^2} - \Gamma_{\rm A_0}^2 \frac{\pd^2 v_r}{\pd z^2} = - \frac{1}{\rho_0}  \frac{\pd^2 p_{{\rm T}}}{\pd t \pd r}. \label{eq:ptvr}
\end{equation}
Note that Equation~(\ref{eq:pt11}) is only valid in the regions with homogeneous densities, hence it cannot be applied in the transversely inhomogeneous transitional layer. Now, we write all perturbations proportional to $\exp \left( - i \omega t + i m \varphi \right)$, where $m$ is an integer representing the azimuthal wavenumber. In the absence of magnetic twist, both positive and negative values of $m$ are equivalent, so hereafter we restrict ourselves to positive values of $m$. For kink oscillations, $m=1$. Equation~(\ref{eq:pt11}) becomes
\begin{equation}
  \frac{\pd^2 p_{{\rm T}}}{\pd z^2} + \frac{1}{r} \frac{\pd}{\pd r} \left( r  \frac{\pd p_{{\rm T}}}{\pd r} \right) + \left( \frac{\omega^2}{\Gamma_{\rm A_0}^2} - \frac{m^2}{r^2} \right) p_{{\rm T}} = 0.  \label{eq:pt12}
\end{equation}

Since $\rho_{\rm p} $, $\rho_{\rm e}$, and $\rho_{\rm c}$ are uniform in their respective regions, the corresponding Cowling's diffusivities, $\etacf$, $\etace$, and $\etacc$, respectively, are also uniform. Both the corona and the evacuated region are fully ionized and much less dense than the prominence plasma, so we have $\etacc \ll \etacf$ and $\etace \ll \etacf$. For the sake of simplicity, we set $\etace = \etacc = 0$, and the effect of Cowling's diffusion is only considered in the prominence region. If considered, Cowling's diffusion in the evacuated and coronal regions would have a very minor influence since Cowling's diffusivities in these regions are much smaller than that in the prominence plasma.

\subsection{Mathematical Method}

 For a fully ionized plasma, the general investigation of the ideal transverse MHD oscillations supported by our equilibrium was performed by \citet{diaz2002} in the case $l/a = 0$ and for the prominence thread centered within the magnetic tube. These authors obtained the oscillatory frequencies and eigenfunctions for arbitrary values of $L$, $L_{\rm p}$, and $a$. Here, we could follow a treatment similar to that of \citet{diaz2002}, but this requires a significant mathematical effort beyond the purpose of the present investigation. Instead, we consider the much simpler approach introduced by \citet{dymovaruderman}, who studied the same configuration but in the thin tube (TT) limit, i.e., for $a/L \ll 1$ and $a/L_{\rm p} \ll 1$. To check the validity of this approximation in the context of prominence thread oscillations, we take into account the values of $a$ and $L_{\rm p}$ reported from the observations (see Sec.~\ref{sec:equi}) and assume $L \sim 10^5$~km. We obtain $a/L_{\rm p}$ and $a/L$ in the ranges $2\times 10^{-3} \lesssim a/L_{\rm p} \lesssim 0.1$ and $5\times 10^{-4} \lesssim a/L \lesssim 3\times 10^{-3}$, meaning that the TT approximation is justified in prominence fine structures. As shown by \citet{hinode} and \citet{diazperiods}, the method of \citet{dymovaruderman}  shows an excellent agreement with that of  \citet{diaz2002} when realistic values of $a/L_{\rm p}$ and $a/L$ are taken into account.

Following \citet{dymovaruderman}, we can perform a different scaling of Equation~(\ref{eq:pt12}) inside the tube and in the corona. For perturbations inside the tube, the characteristic scale in the $r$-direction is $a$, while the characteristic scale in the $z$-direction is $L$. Since $a/L \ll 1$, the term with the longitudinal derivative and the term proportional to $\omega^2$ are much smaller than the other terms. In such a case, Equation~(\ref{eq:pt12}) inside the tube reduces to
\begin{equation}
 \frac{\pd}{\pd r} \left( r  \frac{\pd p_{{\rm Ti}}}{\pd r} \right) - \frac{m^2}{r^2} p_{{\rm Ti}} \approx 0, \label{eq:ptin}
\end{equation}
with i $=$ p or e. The solution of Equation~(\ref{eq:ptin}) for regular perturbations at $r=0$ is
\begin{equation}
 p_{\rm Ti} \approx A_{\rm i} \left( z \right) \left( \frac{r}{a} \right)^m,
\end{equation}
where  $A_{\rm i} \left( z \right)$ is an arbitrary function of $z$.

On the other hand, the characteristic scale of perturbations outside the tube, i.e., in the corona, is $L$ in both the $r$- and $z$-directions, so that no terms can be neglected in Equation~(\ref{eq:pt12}). However, we can express the total pressure perturbation in the corona as $p_{\rm Tc} = A_{\rm c} \left( z \right) F \left( r \right)$ and use the technique of separation of variables to obtain the following expressions,
\begin{equation}
 \frac{{\rm d}^2 F}{{\rm d} r^2} + \frac{1}{r} \frac{{\rm d} F}{{\rm d} r} - \left( k^2_n + \frac{m^2}{r^2}  \right) F = 0, \label{eq:fapp}
\end{equation}
and
\begin{equation}
 \frac{{\rm d}^2 A_{\rm c}}{{\rm d} z^2} + \frac{\omega^2}{\vac^2} A_{\rm c}= - k^2_n A_{\rm c}, \quad \textrm{with} \quad A_{\rm c}= 0 \quad \textrm{at} \quad z = z^\pm_{\rm wall}, \label{eq:acapp}
\end{equation}
where $k_n$ is a separation constant, with $n$ an integer accounting for the different radial harmonics. In Equation~(\ref{eq:acapp}) we have taken into account that Cowling's diffusion is neglected in the corona, so $\Gamma_{\rm Ac}^2 = \vac^2$. Equation~(\ref{eq:fapp}) is the modified Bessel Equation. Here, we only consider trapped modes and assume $k^2_n > 0$. This last condition may not be satisfied for high harmonics, i.e., large values of $n$,  but we need not worry about this issue since here we focus our investigation on the fundamental mode, which is non-leaky in the present configuration. Then, the solution of  Equation~(\ref{eq:fapp}) is $F \left(r \right) = K_m \left(k_n r \right)$, with $K_m$ the modified Bessel function of the second kind. An asymptotic expansion near the tube boundary \citep[e.g.,][]{abra} allows us to express the total pressure perturbation in the corona as
\begin{equation}
 p_{\rm Tc} \approx A_{\rm c} \left( z \right) \left( \frac{a}{r} \right)^m. \label{eq:pout}
\end{equation}

Next, our method follows closely that of \citet{dymovaruderman,dymovaruderman2}. For the sake of simplicity, we omit here the details, which are given in Appendix~\ref{appendix}. After considering appropriate boundary conditions for the solutions of Equations~(\ref{eq:ptin}) and (\ref{eq:pout}), the dispersion relation for kink oscillations damped by resonant absorption and Cowling's diffusion is obtained.

\section{DISPERSION RELATION AND APPROXIMATIONS}
\label{sec:disper}

The general dispersion relation for kink oscillations is
\begin{eqnarray}
&& \frac{\tilde{c}_{k \rm p} h \cos \left( \frac{\omega}{\tilde{c}_{k \rm p} h} \frac{L_{\rm p}}{2} \right) \cos \left( \frac{\omega}{c_{k \rm e}} L_{\rm e}^-  \right) -  c_{k \rm e} \sin \left( \frac{\omega}{\tilde{c}_{k \rm p} h} \frac{L_{\rm p}}{2} \right) \sin \left( \frac{\omega}{c_{k \rm e}} L_{\rm e}^-  \right) }{\tilde{c}_{k \rm p} h \sin \left( \frac{\omega}{\tilde{c}_{k \rm p} h} \frac{L_{\rm p}}{2} \right) \cos \left( \frac{\omega}{c_{k \rm e}} L_{\rm e}^-  \right) +  c_{k \rm e} \cos \left( \frac{\omega}{\tilde{c}_{k \rm p}h} \frac{L_{\rm p}}{2} \right) \sin \left( \frac{\omega}{c_{k \rm e}} L_{\rm e}^-  \right)  } \nonumber \\
&+& \frac{\tilde{c}_{k \rm p} h \cos \left( \frac{\omega}{\tilde{c}_{k \rm p} h} \frac{L_{\rm p}}{2} \right) \cos \left( \frac{\omega}{c_{k \rm e}} L_{\rm e}^+  \right) -  c_{k \rm e} \sin \left( \frac{\omega}{\tilde{c}_{k \rm p} h} \frac{L_{\rm p}}{2} \right) \sin \left( \frac{\omega}{c_{k \rm e}} L_{\rm e}^+  \right) }{\tilde{c}_{k \rm p} h \sin \left( \frac{\omega}{\tilde{c}_{k \rm p} h} \frac{L_{\rm p}}{2} \right) \cos \left( \frac{\omega}{c_{k \rm e}} L_{\rm e}^+ \right) +  c_{k \rm e} \cos \left( \frac{\omega}{\tilde{c}_{k \rm p}h} \frac{L_{\rm p}}{2} \right) \sin \left( \frac{\omega}{c_{k \rm e}} L_{\rm e}^+  \right)  } = 0, \nonumber \\ \label{eq:disperappcomp}
\end{eqnarray}
with 
\begin{equation}
 \tilde{c}_{k \rm p}^2 = \frac{\frac{\rho_{\rm p} \Gamma_{\rm A p}^2+ \rho_{\rm c} \vac^2}{\rho_{\rm p} + \rho_{\rm c}} - i \pi \omega \left( \Gamma_{\rm A p}^2 + \vac^2 \right) \left( \frac{ \rho_{\rm p} \rho_{\rm c}}{\rho_{\rm p} + \rho_{\rm c}} \right) \frac{m/a}{ \omega_{\rm R}   \left| \pd_r \rho_0 \right|_a} }{1 - i \pi \omega \left( \frac{ \rho_{\rm p} \rho_{\rm c}}{\rho_{\rm p} + \rho_{\rm c}} \right) \frac{m/a}{ \omega_{\rm R}   \left| \pd_r \rho_0 \right|_a}}, \label{eq:ckfull1} 
\end{equation}
\begin{equation}
 b^2 = - \frac{i \omega \pi \Gamma_{\rm A p}^2 \vac^2 \left( \frac{ \rho_{\rm p} \rho_{\rm c}}{\rho_{\rm p} + \rho_{\rm c}} \right) \frac{m/a}{ \omega_{\rm R}   \left| \pd_r \rho_0 \right|_a} }{\frac{\rho_{\rm p} \Gamma_{\rm A p}^2+ \rho_{\rm c} \vac^2}{\rho_{\rm p} + \rho_{\rm c}} - i \pi \omega \left( \Gamma_{\rm A p}^2 + \vac^2 \right) \left( \frac{ \rho_{\rm p} \rho_{\rm c}}{\rho_{\rm p} + \rho_{\rm c}} \right) \frac{m/a}{ \omega_{\rm R}   \left| \pd_r \rho_0 \right|_a} }, \label{eq:bfull}
\end{equation}
\begin{equation}
 c_{k \rm p}^2 = \frac{\rho_{\rm p} \vaf^2+ \rho_{\rm c} \vac^2}{\rho_{\rm p} + \rho_{\rm c}}, \qquad c_{k \rm e}^2 = \frac{\rho_{\rm e} \vae^2+ \rho_{\rm c} \vac^2}{\rho_{\rm e} + \rho_{\rm c}}, \label{eq:kinkspeed}
\end{equation}
and $h = \sqrt{ 1 - \frac{b^2}{c_{k \rm p}^2} }$, with $\omega_{\rm R}$ the real part of the frequency and $ \left| \pd_r \rho_0 \right|_a$ the radial derivative of the density profile evaluated at the resonance position, which has been approximated by the thread mean radius, $a$. The quantity $\tilde{c}_{k \rm p}$ is here called the modified kink speed, which takes into account both the effect of Cowling's diffusion (through $\Gamma_{\rm A p}^2$) and the effect of resonant absorption in the thin boundary (TB) approach. Extensive details are given in Appendix~\ref{appendix}. If the terms related to resonant absorption are omitted, one has $b^2 = 0$ and $\tilde{c}_{k \rm p}^2$ becomes
\begin{equation}
 \tilde{c}_{k \rm p}^2 = \frac{\rho_{\rm p} \Gamma_{\rm A p}^2+ \rho_{\rm c} \vac^2}{\rho_{\rm p}+ \rho_{\rm c}}, \label{eq:cknocapa}
\end{equation}
which reduces to the ideal kink speed, $c_{k \rm p}^2$ (Equation~(\ref{eq:kinkspeed})), when Cowling's diffusion is neglected, i.e., $\Gamma_{\rm A p}^2 = \vaf^2$.

Equation~(\ref{eq:disperappcomp}) is a transcendental equation that has to be solved numerically. Some analytical progress can be performed if the prominence thread is centered within the tube, i.e., $L_{\rm e}^- = L_{\rm e}^+ = \frac{1}{2} \left( L - L_{\rm p} \right)$, and we focus on the fundamental kink mode. This solution corresponds to the mode with the lowest frequency. In such a case, Equation~(\ref{eq:disperappcomp}) can be simplified to
\begin{equation}
 \frac{1}{\tilde{c}_{k \rm p}\sqrt{ 1 - \frac{b^2}{c_{k \rm p}^2} }} \tan \left( \frac{\omega}{\tilde{c}_{k \rm p}\sqrt{ 1 - \frac{b^2}{c_{k \rm p}^2} }} \frac{L_{\rm p}}{2} \right) - \frac{1}{c_{k \rm e}} \cot \left[ \frac{\omega}{c_{k \rm e}}  \left( \frac{L - L_{\rm p}}{2} \right) \right] = 0. \label{eq:disperapp}
\end{equation}
The fundamental kink mode is given by the first root of Equation~(\ref{eq:disperapp}). A first-order Taylor expansion for small arguments of the trigonometric functions of Equation~(\ref{eq:disperapp}) provides us with an approximation to the frequency as
\begin{equation}
 \omega^2 \approx \frac{4}{\left( L- L_{\rm p} \right) L_{\rm p}} \tilde{c}_{k \rm p}^2 \left( 1 - \frac{b^2}{c_{k \rm p}^2} \right). \label{eq:appoxw2}
\end{equation}
We expect Equation~(\ref{eq:appoxw2}) to be valid when both $\omega$ and $L_{\rm p} / L$ are small quantities, so that the arguments of the trigonometric functions of Equation~(\ref{eq:disperapp}) remain small. Note that Equation~(\ref{eq:appoxw2}) fails to represent the kink mode frequency in the limits $L_{\rm p}/L \to 1$ and $L_{\rm p}/L \to 0$, so one should consider intermediate values of $L_{\rm p}/L$ in Equation~(\ref{eq:appoxw2}). The correct expressions for the fundamental kink mode frequency in these limits are
\begin{equation}
 \omega = \frac{\pi}{L} \tilde{c}_{k \rm p}\sqrt{ 1 - \frac{b^2}{c_{k \rm p}^2} }, \qquad \textrm{for} \qquad L_{\rm p}/L  = 1,
\end{equation}
and
\begin{equation}
 \omega = \frac{\pi}{L} c_{k \rm e}, \qquad \textrm{for} \qquad L_{\rm p}/L  = 0.
\end{equation}

We can extract two main results from Equation~(\ref{eq:appoxw2}). First of all, Equation~(\ref{eq:appoxw2}) only depends on the physical properties of the prominence region and the corona through $\tilde{c}_{k \rm p}$, $c_{k \rm p}$, and $b$, and includes no contributions from the evacuated part. And second, the form of Equation~(\ref{eq:appoxw2}) is similar to the approximation of the ideal kink mode frequency in a homogeneous tube, i.e., $\omega^2 \approx k_z^2 c_{k \rm p}^2$, where $k_z$ is the longitudinal wavenumber. Thus, it seems that the main differences between the expression for the homogeneous tube and that for the partially filled tube are that $4/(L-L_{\rm p})L_{\rm p}$ replaces $k_z^2$, and that a redefined kink speed has to be taken into account. This approximation of the frequency is similar to that obtained by \citet{joarder92b} and \citet{oliverslab} for the string (or hybrid) modes of their slab configuration and to that obtained by \citet{diazperiods} in the context of thread seismology using period ratios.  However, we must bear in mind that Equation~(\ref{eq:appoxw2}) is only a first-order approximation to the kink mode frequency.

In the general case, i.e., when the prominence region is allowed to be at any position within the tube, one should consider the dispersion relation given by Equation~(\ref{eq:disperappcomp}). We can follow the same procedure as before and perform a first-order Taylor expansion for small arguments of the trigonometric functions of Equation~(\ref{eq:disperappcomp}). Then, the following approximation for the frequency is obtained,
\begin{equation}
 \omega^2 \approx \frac{4 L}{\left[\left( L- L_{\rm p} \right) L_{\rm p} + 4 L_{\rm e}^- L_{\rm e}^+ \right] L_{\rm p}} \tilde{c}_{k \rm p}^2 \left( 1 - \frac{b^2}{c_{k \rm p}^2} \right). \label{eq:appoxw2gen}
\end{equation}
Note that the only information from the evacuated zones present in Equation~(\ref{eq:appoxw2gen}) is their lengths $L_{\rm e}^-$ and $L_{\rm e}^+$, but no additional physical property of these zones contributes to Equation~(\ref{eq:appoxw2gen}). In the centered case, $L_{\rm e}^- = L_{\rm e}^+ = \frac{1}{2} \left( L - L_{\rm p} \right)$, Equation~(\ref{eq:appoxw2gen}) reverts to Equation~(\ref{eq:appoxw2}). Now, we can consider the limits $L_{\rm e}^- \to 0$ or $L_{\rm e}^+ \to 0$, which correspond to the prominence thread totally displaced toward an end of the tube. In such limits,  Equation~(\ref{eq:appoxw2gen}) becomes
\begin{equation}
 \omega^2 \approx \frac{4 L}{\left( L- L_{\rm p} \right) L_{\rm p}^2} \tilde{c}_{k \rm p}^2 \left( 1 - \frac{b^2}{c_{k \rm p}^2} \right). \label{eq:appoxw2disp}
\end{equation}
The ratio of Equation~(\ref{eq:appoxw2}) to Equation~(\ref{eq:appoxw2disp}) estimates the shift of $\omega^2$ when the thread is displaced from the center toward the end of the tube. Denoting this ratio as $\delta \omega^2$, we obtain $\delta \omega^2 = L_{\rm p}/L$. Since $L_{\rm p}/L <1$, we expect the frequency of the kink mode to increase as the prominence thread is displaced form the central position. Obviously, for $L_{\rm p}/L = 1$ there is no frequency shift because the prominence plasma occupies the whole tube.

\subsection{Period of the fundamental kink mode}

\label{sec:period}

First, we focus our analytical investigation on the period. Here, we neglect the effect of the damping mechanisms since we assume that the kink mode period is only slightly affected by the presence of the damping mechanisms. As a first approximation, we consider Equation~(\ref{eq:appoxw2gen}) with $\tilde{c}_{k \rm p} = c_{k \rm p}$ and $b=0$, so now $\omega$ is a real quantity. We compute the period as $P = 2\pi / \omega$, obtaining
\begin{equation}
P = \frac{\pi}{\vaf} \sqrt{\frac{\rho_{\rm p} + \rho_{\rm c}}{2 \rho_{\rm p}}} \sqrt{\frac{\left[\left( L- L_{\rm p} \right) L_{\rm p} + 4 L_{\rm e}^- L_{\rm e}^+ \right] L_{\rm p}}{L}}. \label{eq:periodgen}
\end{equation}
Since the expression for the period is known, we could compare the theoretical periods with those observed and apply the MHD seismology technique to prominence fine structure oscillations. However, in our case $P$ depends on many parameters of the model, so Equation~(\ref{eq:periodgen}) alone is not very useful from a seismological point of view. If we assume that the prominence thread is located at the center of the magnetic tube, Equation~(\ref{eq:periodgen}) becomes
\begin{equation}
 P \approx \frac{\pi}{\vaf}  \sqrt{\frac{\rho_{\rm p} + \rho_{\rm c}}{2 \rho_{\rm p}}}  \sqrt{\left( L- L_{\rm p} \right) L_{\rm p}}. \label{eq:periodgen2}
\end{equation}
In addition, in the case of prominences one has $\rho_{\rm p} \gg \rho_{\rm c}$, so Equation~(\ref{eq:periodgen2}) can be simplified to
\begin{equation}
 P \approx \frac{\pi}{\sqrt{2}\vaf} \sqrt{\left( L- L_{\rm p} \right) L_{\rm p}}. \label{eq:periodgen3}
\end{equation}
Although the value of $L_{\rm p} $ can be measured from H$\alpha$ observations of filaments \citep[e.g.,][]{lin07,lin09}, we still have two parameters, i.e., the total tube length, $L$, and the prominence Alfv\'en speed, that are both difficult to determine from the observations. Recently, \citet{diazperiods} showed that the ratio of periods of different overtones is a useful quantity to perform seismology of prominence threads, since additional parameters as, e.g., the prominence Alfv\'en speed are dropped from the expressions \citep[for details about the importance of the period ratio for coronal seismology, see the recent review by][]{reviewperiods}. Some additional remarks about the MHD seismology technique are given in Section~\ref{sec:seismology}.

\subsection{Damping by Cowling's diffusion}

\label{sec:dampcowlingana}

Here we study the kink mode damping. Let us consider the case without transverse transitional layer, i.e., $l/a = 0$, so the damping is exclusively due to Cowling's diffusion and the frequency is complex, $\omega = \omega_{\rm R} + i \omega_{\rm I}$, with $\omega_{\rm R}$ and $\omega_{\rm I}$ the real and imaginary parts of the frequency, respectively. Then, $\tilde{c}_{k \rm p}$ is given by Equation~(\ref{eq:cknocapa}) and $b^2 = 0$ since there is no resonant damping for $l/a = 0$. For simplicity, we consider that the prominence thread is located at the central position. Equation~(\ref{eq:appoxw2}) allows us to obtain the real and imaginary parts of the frequency as
\begin{eqnarray}
 \omega &\approx&  \left[ \frac{\rho_{\rm p} \vaf^2+ \rho_{\rm c} \vac^2}{\rho_{\rm p}+ \rho_{\rm c}} - \left( \frac{\rho_{\rm p} \etacf}{\rho_{\rm p} + \rho_{\rm c}} \right)^2 \frac{1}{\left( L- L_{\rm p} \right) L_{\rm p}}\right]^{1/2}  \frac{2}{\sqrt{\left( L- L_{\rm p} \right) L_{\rm p}}}  \nonumber \\ &-& i \left( \frac{\rho_{\rm p} \etacf}{\rho_{\rm p} + \rho_{\rm c}} \right) \frac{2}{\left( L- L_{\rm p} \right) L_{\rm p}}. \label{eq:approxwcowling}
\end{eqnarray}
 By setting the real part of Equation~(\ref{eq:approxwcowling}) equal to zero, we obtain two critical values of $L_{\rm p}/L$, namely
\begin{equation}
 \left( L_{\rm p} / L \right)_{\rm crit}^\pm = \frac{1}{2} \pm \frac{1}{2} \left[1 - \left( \frac{2 \rho_{\rm p} }{\rho_{\rm p} + \rho_{\rm c}} \right) \etactf^2  \right]^{1/2},  \label{eq:critw}
\end{equation}
with $\etactf = \etacf / \vaf L$. Hence, the kink mode only exists for $ \left( L_{\rm p} / L \right)_{\rm crit}^- <  L_{\rm p} / L <  \left( L_{\rm p} / L \right)_{\rm crit}^+$. We cast Equation~(\ref{eq:critw}) for $\rho_{\rm p} / \rho_{\rm c} = 200$ and the extreme case of an almost neutral plasma with $\mutildef = 0.99$, obtaining $\left( L_{\rm p} / L \right)_{\rm crit}^- \approx 10^{-5}$ and $\left( L_{\rm p} / L \right)_{\rm crit}^+ \approx 0.99999$. For smaller values of $\mutildef$, $\left( L_{\rm p} / L \right)_{\rm crit}^-$ decreases and  $\left( L_{\rm p} / L \right)_{\rm crit}^+$ increases. Hence, the presence of these critical values is irrelevant for realistic values of $L_{\rm p}/L$.

For $L_{\rm p}/L$ far from the critical values, one can drop the second term in the real part of Equation~(\ref{eq:approxwcowling}). We compute the damping time, $\td = 1/|\omega_{\rm I}|$, due to Cowling's diffusion as
\begin{equation}
 \td \approx \frac{1}{2} \left( \frac{\rho_{\rm p} + \rho_{\rm c}}{\rho_{\rm p} \etacf} \right) \left( L- L_{\rm p} \right) L_{\rm p},\label{eq:tdkinkcowapp}
\end{equation}
while the ratio of the damping time to the period is
\begin{equation}
\frac{\td}{P} \approx \frac{1}{2\pi} \left( \frac{\rho_{\rm p}+ \rho_{\rm c} }{\rho_{\rm p}} \right)^{1/2} \frac{1}{\etactf} \sqrt{2 \left( 1- \frac{L_{\rm p}}{L} \right) \frac{L_{\rm p}}{L}}. \label{eq:tdpkinkcowapp}
\end{equation}
For $\rho_{\rm p} / \rho_{\rm c} = 200$, $L_{\rm p}/L = 0.1$, and $L = 10^5$~km, Equation~(\ref{eq:tdpkinkcowapp}) gives $\tdp \approx 5 \times 10^3$ for $\mutildef = 0.8$, and $\tdp \approx 150$ for $\mutildef = 0.99$. These results indicate that, as in a homogeneous thread, an almost neutral prominence plasma is needed, i.e., $\mutildef \approx 1$, for the damping due to Cowling's diffusion to be efficient. Although the precise ionization degree is unknown, such large values of $\mutildef $ are probably unrealistic in the context of prominences.

It is straight-forward to extend Equation~(\ref{eq:tdpkinkcowapp}) to the case in which the prominence region is not at the center of the tube, obtaining
\begin{equation}
\frac{\td}{P} \approx \frac{1}{2\pi} \left( \frac{\rho_{\rm p}+ \rho_{\rm c} }{\rho_{\rm p}} \right)^{1/2} \frac{1}{\etactf} \sqrt{2 \left[ \left( 1- \frac{L_{\rm p}}{L} \right) \frac{L_{\rm p}}{L} + 4 \frac{L_{\rm e}^- L_{\rm e}^+}{L^2} \right] \frac{L_{\rm p}}{L}}. \label{eq:tdpkinkcowappgen}
\end{equation}
Performing the limits $L_{\rm e}^- \to 0$ or $L_{\rm e}^+ \to 0$ to this last expression, we can obtain the damping ratio when the thread is totally displaced toward the ends of the tube, namely $\left( \td / P \right)_{\rm end}$. Then, we compare $\left( \td / P \right)_{\rm end}$ with the damping ratio obtained in the centered case from Equation~(\ref{eq:tdpkinkcowapp}), namely $\left( \td / P \right)_{\rm center}$. Thus, $\left( \td / P \right)_{\rm center} / \left( \td / P \right)_{\rm end} = \sqrt{ L_{\rm p} / L} < 1$. Therefore, the minimum value of the damping ratio by Cowling's diffusion takes place when prominence region is located at the center.

\subsection{Damping by resonant absorption}
 
\label{sec:resonantabst}

Here, we study the general case $l/a \ne 0$. The full expressions of $\tilde{c}_{k \rm p}$ and $b^2$ given by Equations~(\ref{eq:ckfull1}) and (\ref{eq:bfull}) are taken into account. Again, we assume that the prominence thread is centered within the magnetic tube.  We use Equation~(\ref{eq:appoxw2}) to provide an expression for the ratio $ \omega_{\rm I} / \omega_{\rm R}$ after neglecting terms of $\mathcal{O} \left( \omega_{\rm I}^2 \right)$ and $\mathcal{O} \left( \omega_{\rm I} L^{-2} \right)$. Thus, we obtain
\begin{equation}
 \frac{\omega_{\rm I} }{\omega_{\rm R}} \approx -\frac{\pi}{8} \frac{\left( \rho_{\rm p} - \rho_{\rm c} \right)^2}{\left( \rho_{\rm p}+ \rho_{\rm c} \right)}  \frac{m/a}{ \left| \pd_r \rho_0 \right|_a} - \left( \frac{\rho_{\rm p}}{ \rho_{\rm p} + \rho_{\rm c} }\right)^{1/2} \frac{\etactf }{\sqrt{2 \left( 1- \frac{L_{\rm p}}{L} \right)\frac{L_{\rm p}}{L}}}. \label{eq:tdpapproxapp}
\end{equation}
The first term on the right-hand side of Equation~(\ref{eq:tdpapproxapp}) is caused by resonant absorption and the second term is due to Cowling's diffusion. The term related to Cowling's diffusion is also present in the case $l/a = 0$. We can compare Equation~(\ref{eq:tdpapproxapp}) with Equation~(28) of \citet{solerRAPI} valid for a homogeneous tube. We see that both equations coincide if the replacement of $k_z$ by $2/\sqrt{(L-L_{\rm p})L_{\rm p}}$ is done in their expression and $\etactc = 0$. Therefore, this fact suggests again that the results for the homogeneous tube can be extended to a partially filled tube by selecting the appropriate value for the longitudinal wavenumber.

Note that the term related to the damping by resonant absorption in Equation~(\ref{eq:tdpapproxapp})  takes the same form as in a homogeneous tube and does not depend on $L_{\rm p}/L$. This result does not mean that the real and imaginary parts of the frequency do not depend on $L_{\rm p}/L$, but both quantities are affected in the same way so that their ratio remains unaffected. This important result is consistent with the conclusions of \citet{andries2005} and \citet{arregui05}, who found that the kink mode frequency of a longitudinally stratified tube is the same as that obtained for a homogeneous tube with density $\rho_{\rm mean}$, with $\rho_{\rm mean}$ the mean density weighted with the wave energy. Since in our equilibrium the transversely transitional layer is only present in the dense part of the tube and both the evacuated zone and the corona have the same density, resonant absorption only takes place in the dense part of the tube. Therefore, the mean density of the part of the tube where resonant absorption takes place is, obviously,  $\rho_{\rm mean} = \rho_{\rm p}$. Hence, according to \citet{andries2005} and \citet{arregui05}, the kink mode damping ratio in our case must be the same as that of a homogeneous tube with density $\rho_{\rm p}$, as our results indicate. This conclusion is also equivalent to that obtained by \citet{dymovaruderman2}, who showed that the damping ratio in a longitudinally inhomogeneous tube in the TT and TB approximations does not depend on the particular form of the longitudinal density profile if the density contrast between the internal and external plasmas is constant.

By assuming a sinusoidal density variation in the transitional layer, the expression for $\tdp$ according to Equation~(\ref{eq:tdpapproxapp}) is
\begin{equation}
 \frac{\td}{P} \approx \frac{2}{\pi} \left[ m  \frac{l}{a} \frac{\rho_{\rm p} -\rho_{\rm c} }{\rho_{\rm p}+ \rho_{\rm c} }  +  \etactf \left( \frac{\rho_{\rm p}}{\rho_{\rm p}+ \rho_{\rm c} } \right)^{1/2} \frac{4}{\sqrt{2 \left( 1- \frac{L_{\rm p}}{L} \right) \frac{L_{\rm p}}{L}}} \right]^{-1}, \label{eq:appdampingratiotot}
\end{equation}
which is equivalent to Equation~(29) of \citet{solerRAPI} for $\etactc = 0$ and $k_z = 2/\sqrt{(L-L_{\rm p})L_{\rm p}}$. To perform a simple application, we compute $\tdp$ from Equation~(\ref{eq:appdampingratiotot}) in the case $m=1$, $L_{\rm p} / L = 0.1$, $L = 10^7$~m, and $l/a = 0.2$, resulting in $\tdp \approx 3.18$ for a fully ionized thread ($\mut_{\rm p} = 0.5$), and  $\tdp \approx 3.16$ for an almost neutral thread ($\mut_{\rm p} = 0.95$). We note that the obtained damping times are consistent with the observations. Moreover, as obtained by \citet{solerRAPI}, the contribution of resonant absorption to the damping is much more important than that of Cowling's diffusion, so the ratio $\tdp$ depends only very slightly on the ionization degree. Then, the second term on the right-hand side of Equation~(\ref{eq:appdampingratiotot}) can be neglected and Equation~(\ref{eq:appdampingratiotot}) becomes
\begin{equation}
  \frac{\td}{P} \approx \frac{2}{\pi} \frac{1}{m} \frac{1}{ l/a}\frac{\rho_{\rm p} +\rho_{\rm c} }{\rho_{\rm p}- \rho_{\rm c} },  \label{eq:appdampingratiotot2}
\end{equation}
which coincides with the expressions provided by \citet{rudermanroberts02} and \citet{goossens02} in the case of longitudinally homogeneous coronal loops. Equation~(\ref{eq:appdampingratiotot2}) can be further simplified by restricting ourselves to the kink mode ($m=1$) and for $\rho_{\rm p} \gg \rho_{\rm c}$, hence
\begin{equation}
 \frac{\td}{P} \approx \frac{2}{\pi} \frac{1}{ l/a},  \label{eq:tdsimp}
\end{equation}
meaning that the transverse inhomogeneity spatial-scale can be estimated if both the period and damping time are measured from the observations. By combining Equations~(\ref{eq:periodgen3}) and (\ref{eq:tdsimp}), we get the expression of the damping time by resonant absorption as
\begin{equation}
 \td \approx  \frac{\sqrt{2}}{\vaf} \frac{1}{l/a} \sqrt{\left( L - L_{\rm p}  \right) L_{\rm p}}. \label{eq:tdressapp}
\end{equation}

Finally, we can compute the damping ratio when the prominence thread is displaced from the center of the tube. In such a case, only the term related to Cowling's diffusion is modified in the way shown in Section~\ref{sec:dampcowlingana}, while the term related to the resonant damping is not affected at all. The damping time by resonant absorption in the general case is
\begin{equation}
  \td \approx  \frac{\sqrt{2}}{\vaf} \frac{1}{l/a} \sqrt{\frac{\left[\left( L- L_{\rm p} \right) L_{\rm p} + 4 L_{\rm e}^- L_{\rm e}^+ \right] L_{\rm p}}{L}}, \label{eq:tdressappgen}
\end{equation}
which has the same dependence on $L_{\rm e}^-$ and $L_{\rm e}^+$ as the period (Equation~(\ref{eq:periodgen})). Since the resonant damping dominates over Cowling's diffusion, we anticipate by means of this analytical estimations that the damping ratio is almost unaffected by the position of the prominence region within the fine structure.

\section{NUMERICAL RESULTS}
\label{sec:results}

\subsection{Centered prominence thread}
\label{sec:symmetric}

In this section, we assume that the prominence region is centered within the tube, i.e., $L_{\rm e}^- = L_{\rm e}^+ = \frac{1}{2} \left( L - L_{\rm p} \right)$. We numerically solve the dispersion relation (Equation~(\ref{eq:disperapp})) by means of standard methods and obtain the frequency of the fundamental mode. We study the dependence of the results on $L_{\rm p}/L$.

\begin{figure}[!tp]
\centering
 \epsscale{0.49}
 \plotone{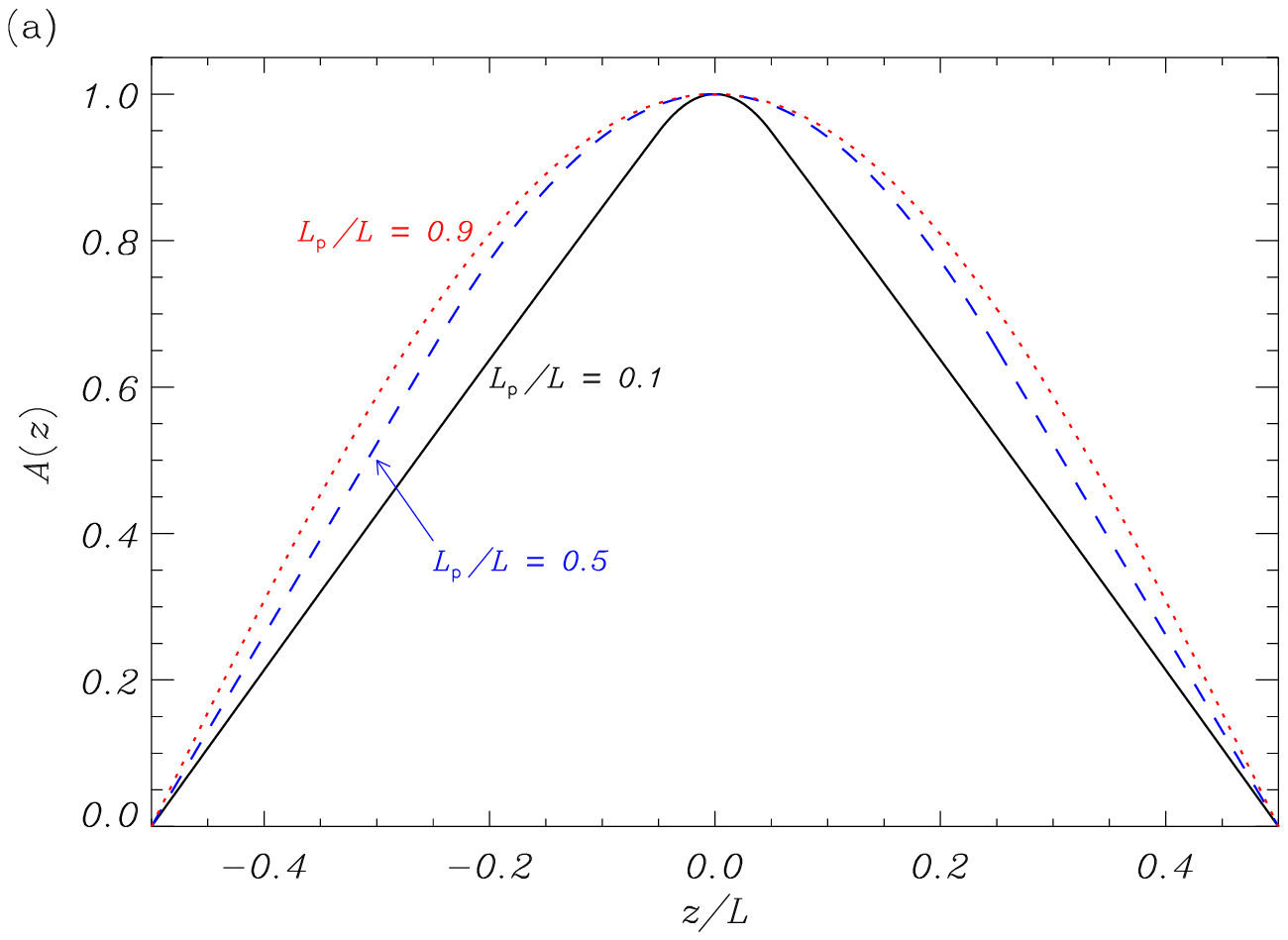}
 \plotone{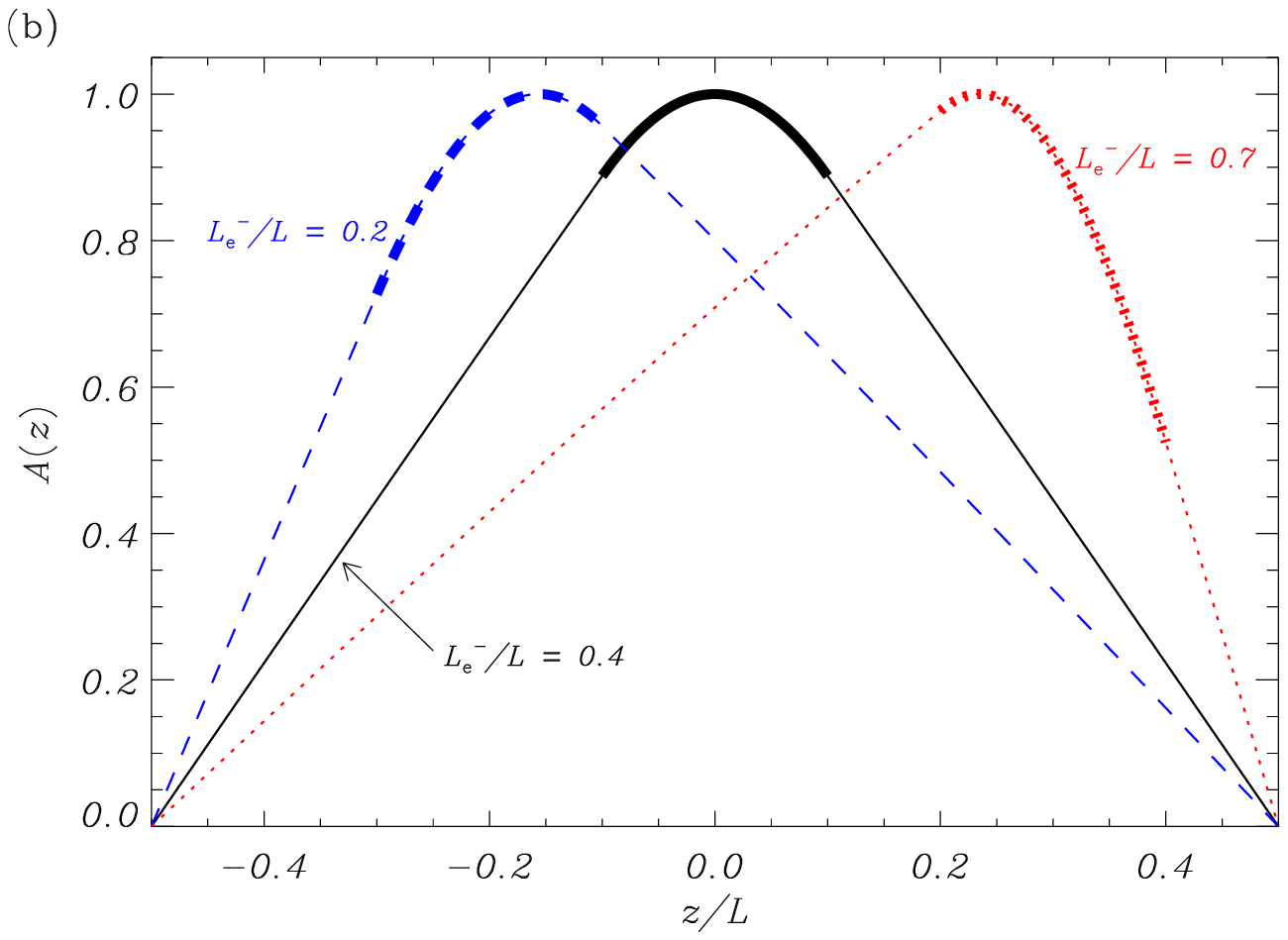}
\caption{$A(z)$ function (in arbitrary units) corresponding to the fundamental kink mode. (a) Results for $L_{\rm p}/L=$~0.1 (solid), 0.5 (dashed), and 0.9 (dotted) when the prominence thread is located at the central part of the magnetic tube. (b) Results with $L_{\rm p}/L=$~0.2 for $L_{\rm e}^-/L=$~0.2 (dashed), 0.4 (solid), and 0.7 (dotted). The thick part of the lines in panel (b) denotes the position of the prominence thread.  \label{fig:eigenaz}}
\end{figure}

We plot in Figure~\ref{fig:eigenaz}a the $A(z)$ function corresponding to the fundamental even kink mode for different values of $L_{\rm p}/L$. The $A(z)$ function gives the dependence of the transverse displacement in the longitudinal direction at $r=a$. We see that $A(z)$ is mainly confined within the prominence part of the flux tube and satisfies the line-tying condition at $z=z^\pm_{\rm wall} = \pm L/2$. For a homogeneous prominence tube, i.e., $L_{\rm p} / L = 1$, the $A(z)$ function becomes proportional to $\cos \left( \frac{\pi}{L}  z  \right)$.

\subsubsection{Case without transverse transitional layer ($l/a=0$)}

\begin{figure}[!tp]
\centering
 \epsscale{0.49}
 \plotone{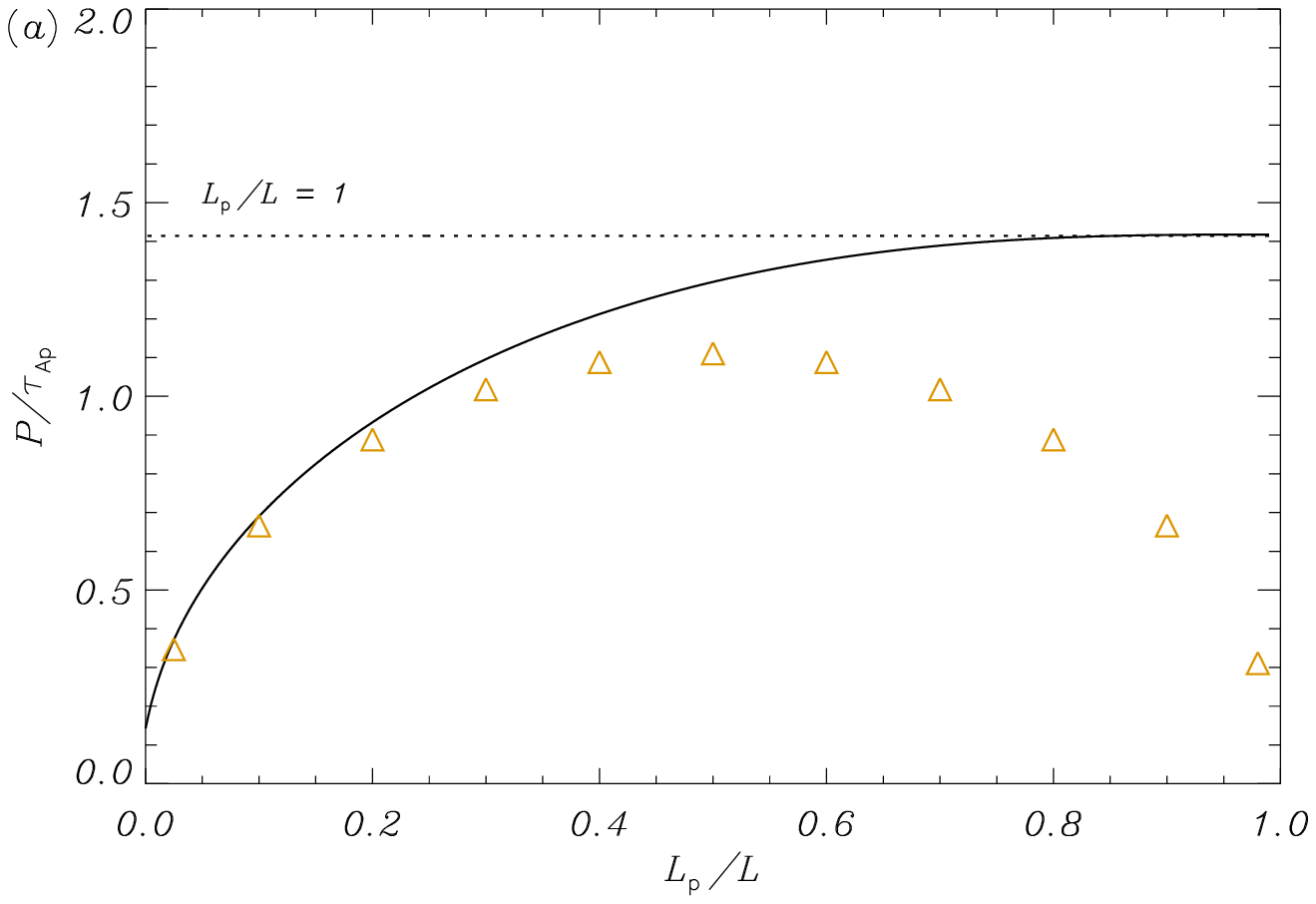}
 \plotone{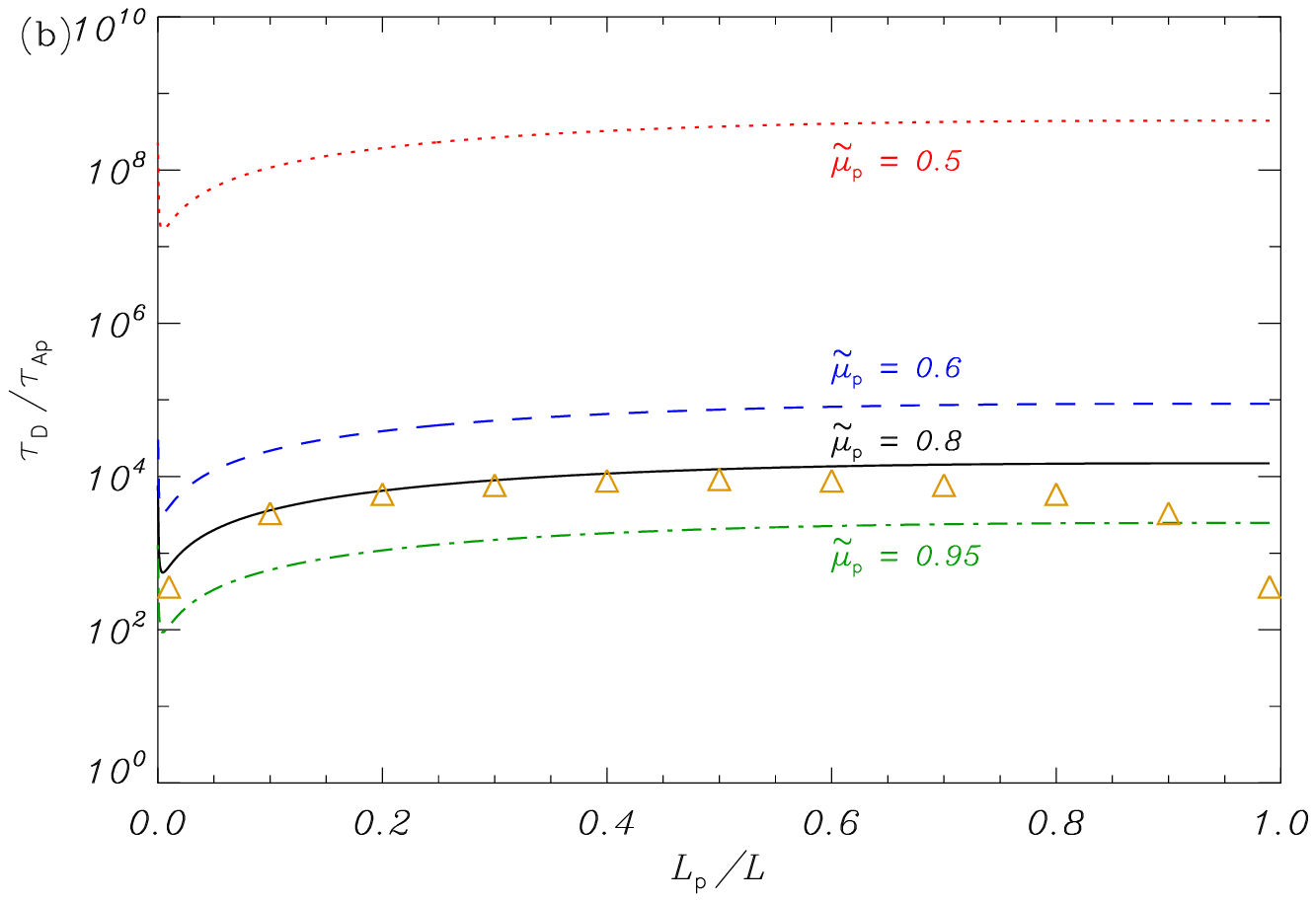}
 \plotone{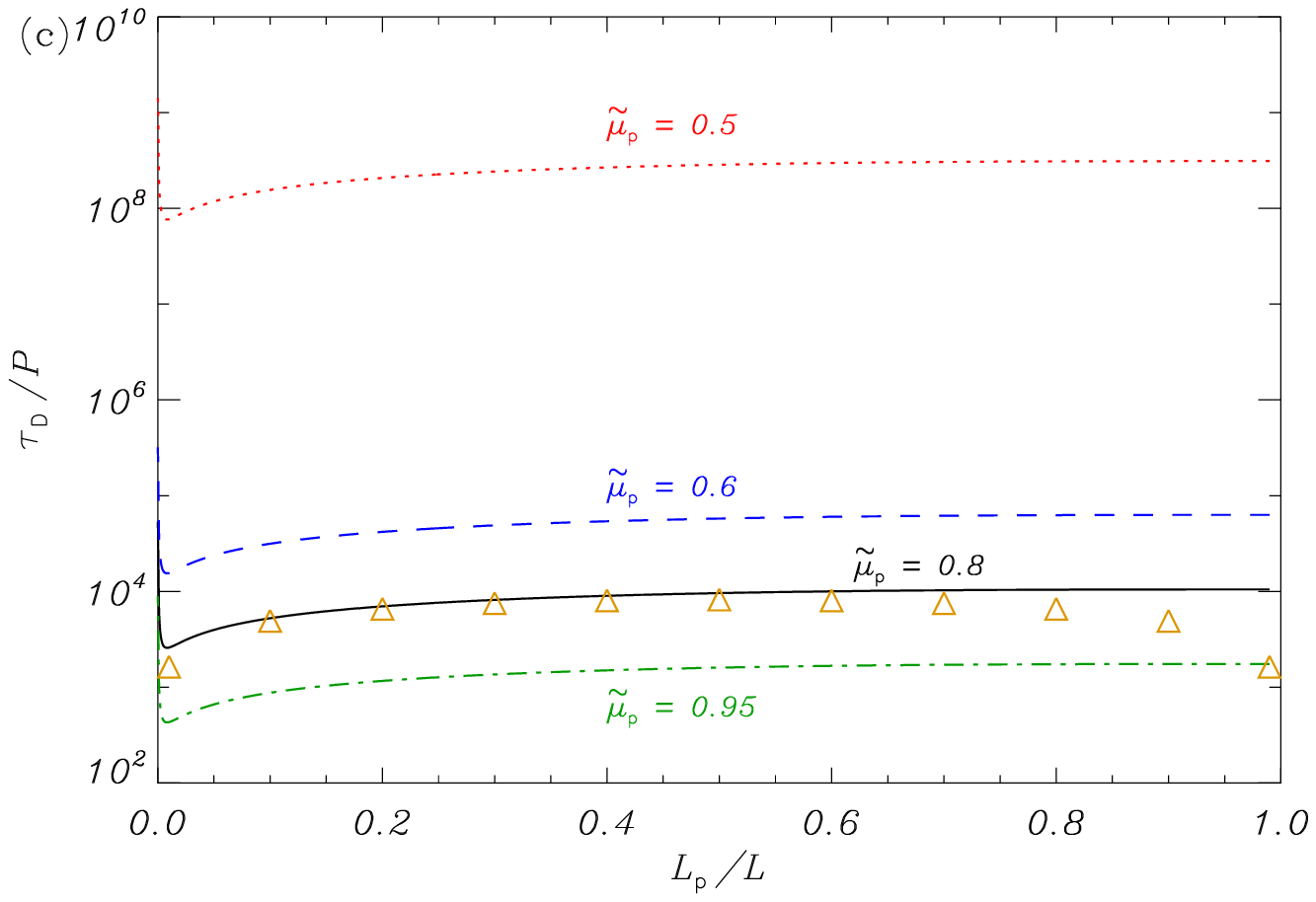}
\caption{Results in the case without transverse transitional layer and for the prominence thread located at the central part of the magnetic tube. (a) Period, $P$, of the fundamental kink mode in units of the internal Alfv\'en travel time, $\tau_{\rm A p}$, as a function of  $L_{\rm p}/L$. The horizontal dotted line corresponds to the period of the kink mode in a homogeneous prominence cylinder. The symbols are the approximation given by Equation~(\ref{eq:periodgen3}). (b) Damping time, $\td$, in units of the internal Alfv\'en travel time, $\tau_{\rm A p}$, as a function of  $L_{\rm p}/L$. The different lines denote $\mutildef=$~0.5 (dotted), 0.6 (dashed), 0.8 (solid) and 0.95 (dash-dotted). The symbols are the approximation given by Equation~(\ref{eq:tdkinkcowapp}) for $\mutildef=$~0.8. (c) $\tdp$ versus $L_{\rm p}/L$. The line styles have the same meaning as in panel (b), and the symbols are the approximation given by Equation~(\ref{eq:tdpkinkcowapp}).  \label{fig:ressapp}}
\end{figure}

First, we take into account the case without transverse transitional layer, i.e., $l/a = 0$, and so we study the kink mode damping  due to Cowling's diffusion exclusively. Figure~\ref{fig:ressapp}a displays the period, $P$, as a function of $L_{\rm p}/L$ for different values of the ionization degree in the prominence region, whereas Figure~\ref{fig:ressapp}b shows the corresponding values of the damping time, $\td$.  Both values are given in dimensionless form with respect to the internal Alfv\'en travel time, $\tau_{\rm A p} = L/\vaf$. We see that $P/\tau_{\rm A p}$ increases as $L_{\rm p}/L$ increases, and tends to the value for a homogeneous prominence cylinder when $L_{\rm p}/L \to 1$. In addition, we find that $P$ is independent of the ionization degree. On the contrary, $\td$ is strongly dependent on the ionization degree, as expected. $\td/\tau_{\rm A p}$ slightly increases as $L_{\rm p}/L$ becomes larger.  We see that the analytical expressions for the period (Equation~(\ref{eq:periodgen3})) and the damping time (Equation~(\ref{eq:tdkinkcowapp})) are in agreement with the solution of the full dispersion relation for realistic, small values of $L_{\rm p}/L$, i.e., $L_{\rm p}/L \lesssim 0.4$, whereas the approximate expressions diverge from the actual solution when the prominence region occupies most of the magnetic tube. As commented before, high-resolution observations suggest that the parameter $L_{\rm p}/L$ is small in the fine structures of prominences. On the other hand, Figure~\ref{fig:ressapp}c displays $\tdp$ versus $L_{\rm p}/L$. The numerical solution of the dispersion relation shows little dependence on $L_{\rm p}/L$, while the analytical approximation (Equation~(\ref{eq:tdpkinkcowapp})) diverges from the numerical value in the limit of large $L_{\rm p}/L$.  Given the large values of $\tdp$ obtained, we can conclude that the efficiency of the damping due to Cowling's diffusion in a partially filled flux tube does not improve with respect to the longitudinally homogeneous tube case of \citet{solerneutrals,solerRAPI}.

\begin{figure}[!tp]
\centering
 \epsscale{0.49}
 \plotone{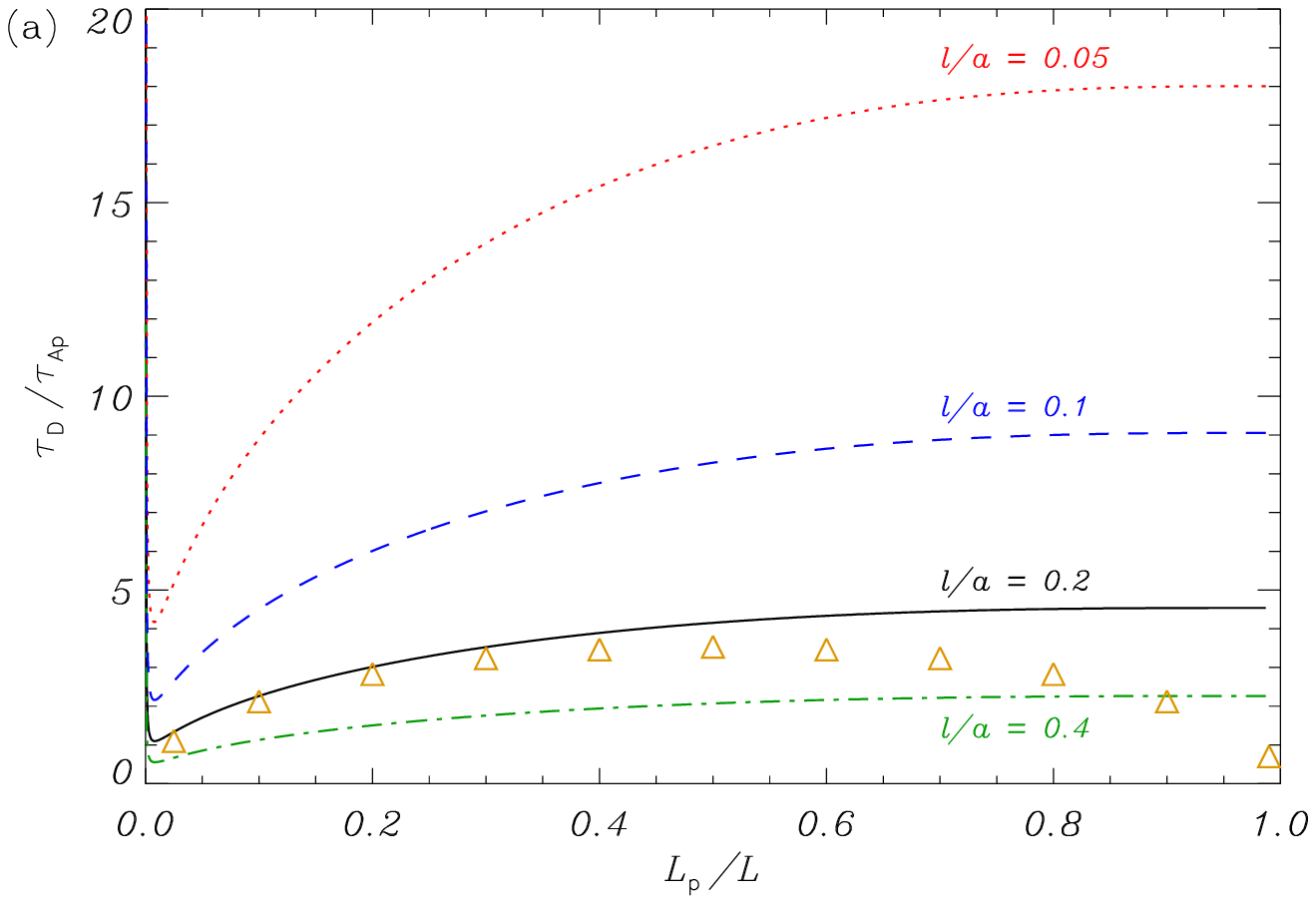}
 \plotone{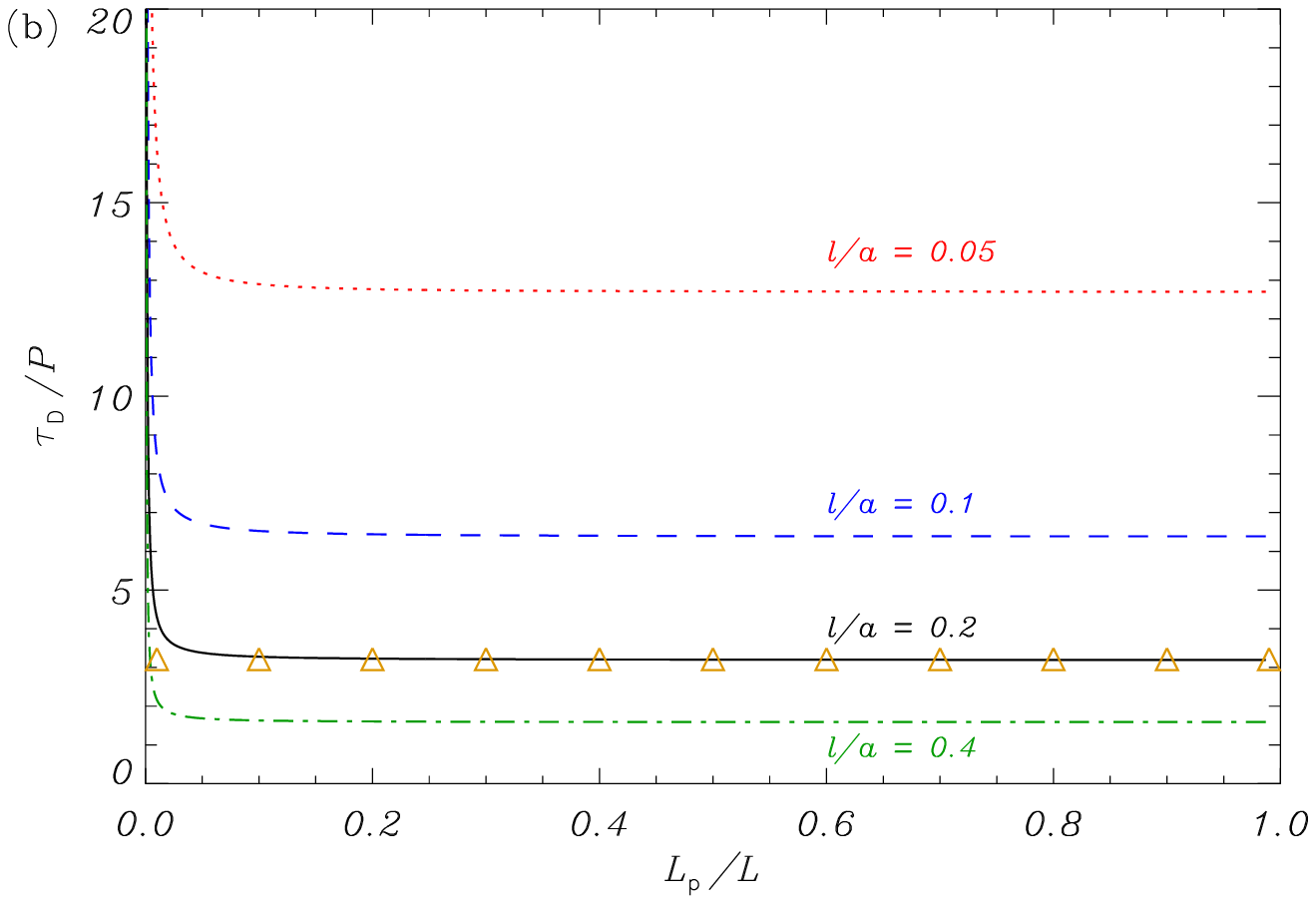}
\caption{Results in the case with a transverse transitional layer and for the prominence thread located at the central part of the magnetic tube. (a) $\td$, in units of the internal Alfv\'en travel time, $\tau_{\rm A p}$, and (b) $\tdp$ as a functions of  $L_{\rm p}/L$. The different lines in both panels denote $l/a=$~0.05 (dotted), 0.1 (dashed), 0.2 (solid) and 0.4 (dash-dotted). The symbols in panels (a) and (b) are the approximations given by Equations~(\ref{eq:tdressapp}) and (\ref{eq:appdampingratiotot2}), respectively, with $l/a=$~0.2.   \label{fig:resapppp}}
\end{figure}

\subsubsection{Case with a transverse transitional layer ($l/a \ne 0$)}

Now, we take the case $l/a \neq 0$ into account. The kink mode is damped by resonant absorption in the transverse transitional layer. We have computed both the period and the damping time of the fundamental kink mode as a function of the different parameters, namely $\mutildef$, $l/a$, and $L_{\rm p}/L$. Regarding the period, we find that both its value and its dependence on 
$L_{\rm p}/L$  are the same plotted in Figure~\ref{fig:ressapp}a because the period is almost independent of $\mutildef$ and $l/a$. For the sake of simplicity, we do not repeat this Figure again and refer to Figure~\ref{fig:ressapp}a. Therefore, the presence of the transverse transitional layer in the TB approximation does not modify the period of kink oscillations with respect to the case $l/a=0$. The period could be slightly affected if thick layers, i.e, $l/a > 1$, are considered and the resistive equations are solved numerically instead of assuming the TB approach \citep[see, e.g.,][]{tom}. This issue will be further addressed by \citet{arregui2d}.

Figure~\ref{fig:resapppp}a shows $\td/\tau_{\rm A p}$ versus $L_{\rm p}/L$ for different values of $l/a$. These computations correspond to an ionization degree $\mutildef = 0.8$, but equivalent computations for other values of $\mutildef$ provide almost identical results because the effect of Cowling's diffusion is negligible in comparison to that of resonant absorption.  As expected, the value of the damping time decreases with $l/a$. The approximate value of $\td$ given by Equation~(\ref{eq:tdressapp}) is in good agreement with the full solution for $L_{\rm p}/L \lesssim 0.4$, as happens for the period. In order to assess the efficiency of the resonant damping, Figure~\ref{fig:resapppp}b displays the corresponding values of $\tdp$. In comparison to the damping ratio by Cowling's diffusion (see Fig.~\ref{fig:ressapp}c), much smaller values of $\tdp$ are now obtained. As predicted analytically by Equation~(\ref{eq:appdampingratiotot2}), $\tdp$ is almost independent of $L_{\rm p}/L$. By comparing Figures~\ref{fig:ressapp}a and \ref{fig:resapppp}a, we see that both the period and the damping time have a very similar dependence on $L_{\rm p}/L$, so the dependence on $L_{\rm p}/L$ is canceled when the damping ratio is computed. In this case, a very good agreement between the numerical result and the analytical approximation (Equation~(\ref{eq:appdampingratiotot2})) is found even for large values of $L_{\rm p}/L$.

\subsection{Effect of the position of the prominence thread within the magnetic tube}
\label{sec:nonsymmetric}

\begin{figure}[!tp]
\centering
 \epsscale{0.49}
 \plotone{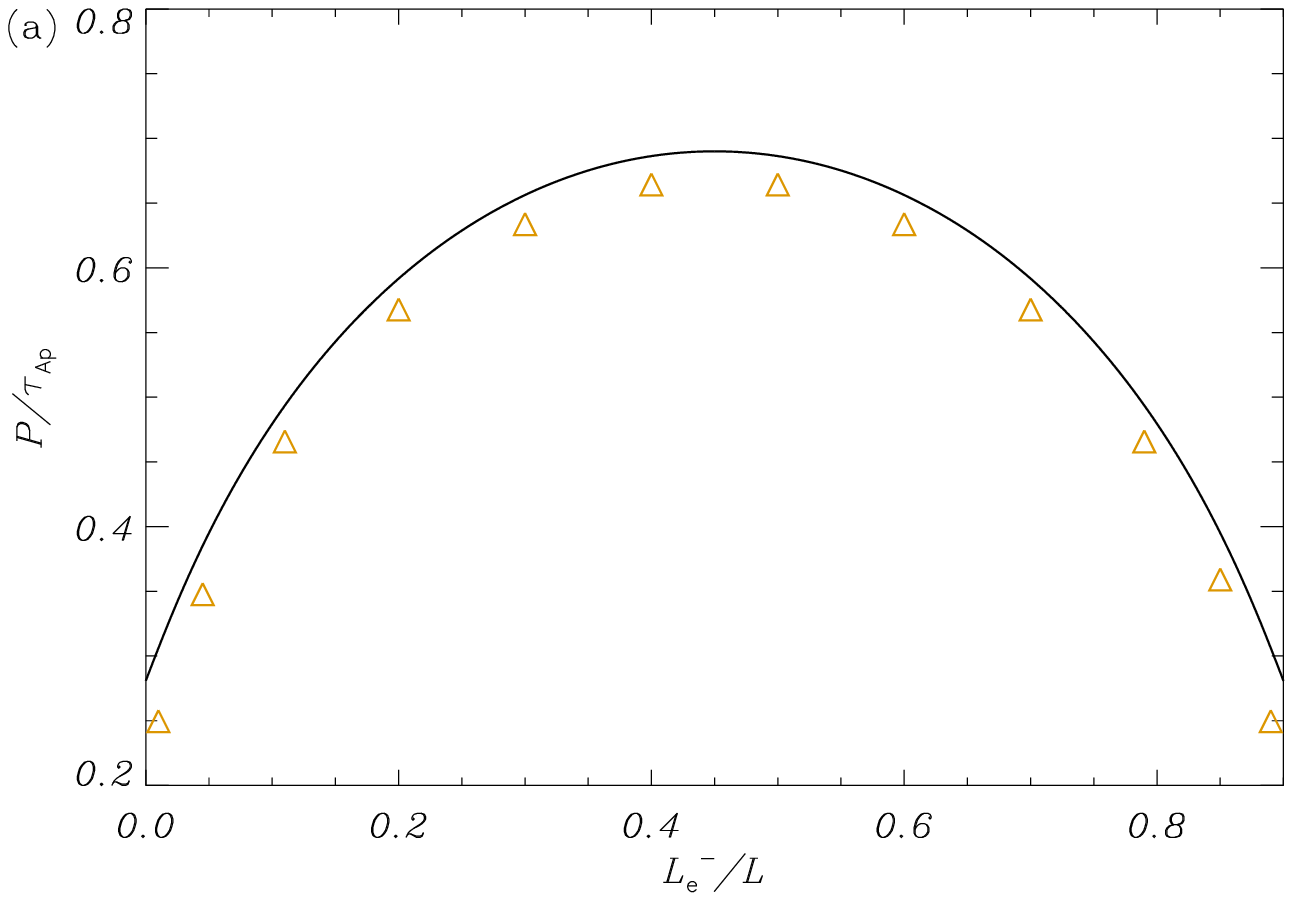}
 \plotone{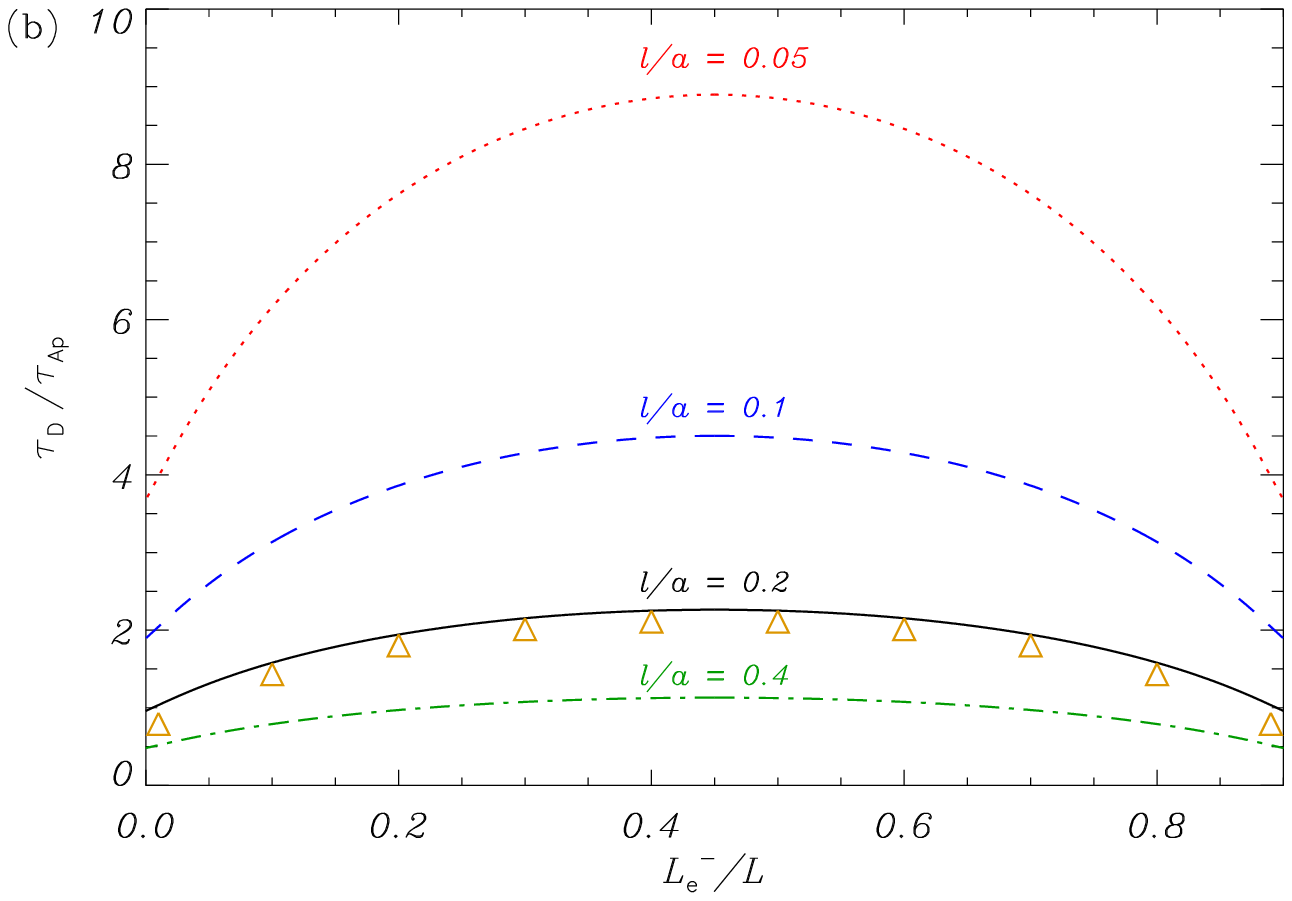}
 \plotone{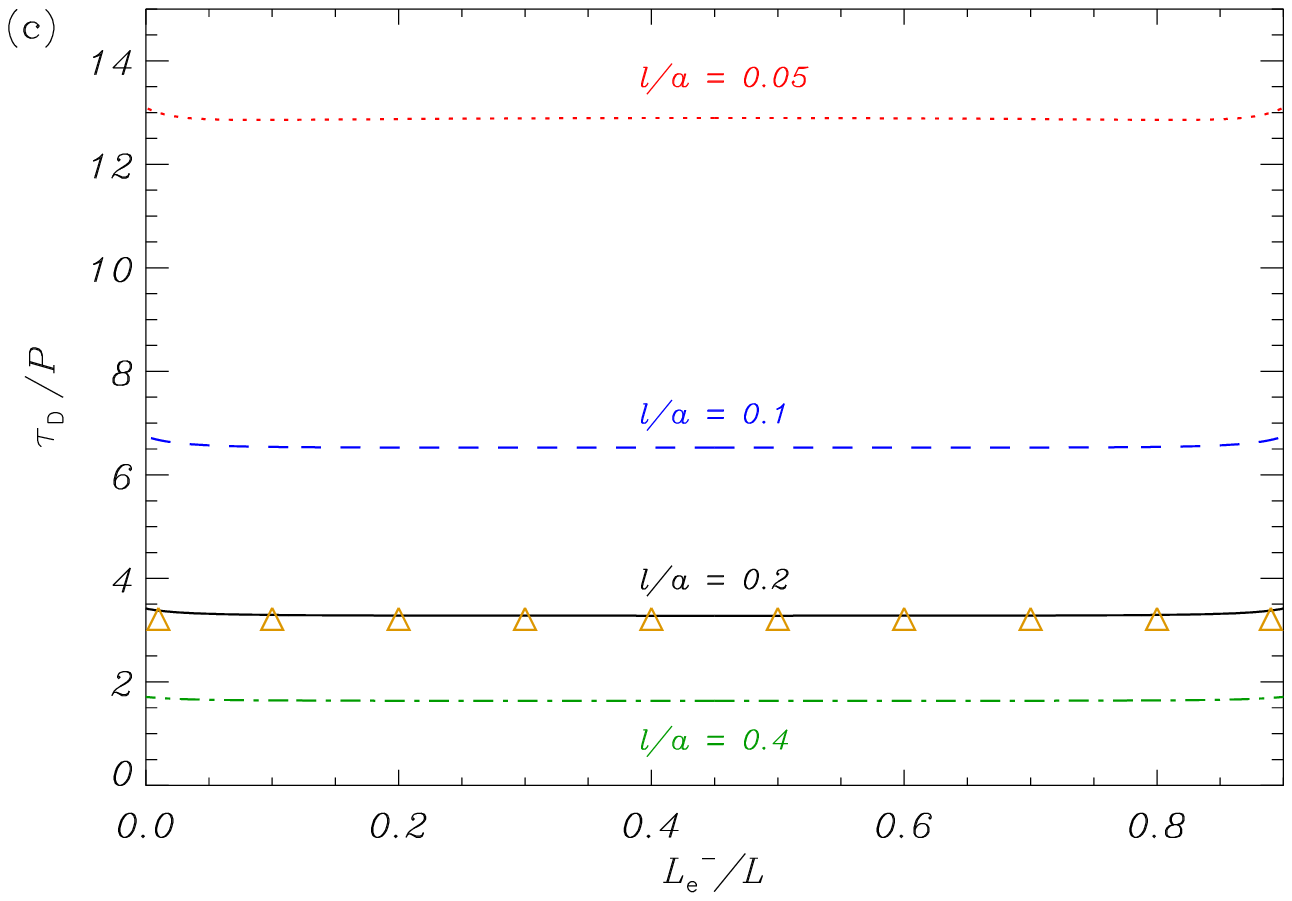}
\caption{(a) Period, $P$, of the fundamental kink mode in units of the internal Alfv\'en travel time, $\tau_{\rm A p}$, as a function of  $L_{\rm e}^-/L$. The symbols are the approximation given by Equation~(\ref{eq:periodgen}). (b) Damping time, $\td$, in units of the internal Alfv\'en travel time, $\tau_{\rm A p}$, as a function of  $L_{\rm e}^-/L$. The different lines denote  $l/a=$~0.05 (dotted), 0.1 (dashed), 0.2 (solid) and 0.4 (dash-dotted). The symbols are the approximation given by Equation~(\ref{eq:tdressappgen}) with $l/a=$~0.2. (c) $\tdp$ versus $L_{\rm e}^-/L$. The line styles have the same meaning as in panel (b), and the symbols are the approximation given by Equation~(\ref{eq:appdampingratiotot2}) with $l/a=$~0.2. In all computations, $L_{\rm p} / L = 0.1$. \label{fig:displaced1}}
\end{figure}

In this Section we study the effect of the position of the prominence region within the magnetic flux tube. The results of the previous Section~\ref{sec:symmetric} correspond to the case in which the thread is located at the center of the cylinder. Here, we allow the dense region to be displaced from the center of the tube. Hence, we must consider the general dispersion relation given by Equation~(\ref{eq:disperappcomp}). The dispersion relation is solved numerically for the lowest frequency solution, equivalent to the fundamental kink mode of the centered case. 

We display in Figure~\ref{fig:eigenaz}b the $A(z)$ function for different values of $L_{\rm e}^- / L$ for $L_{\rm p} / L = 0.2$ and $\mutilde_{\rm p} = 0.8$. Since the oscillation is dominated by the prominence physical properties, we see that the maximum of $A(z)$ is always in the prominence region (denoted by the thick part of the lines in Fig.~\ref{fig:eigenaz}b), regardless of its location within the flux tube.

Next, we plot the period (Fig.~\ref{fig:displaced1}a) and the damping time (Fig.~\ref{fig:displaced1}b) as functions of $L_{\rm e}^-/L$ for $L_{\rm p} / L = 0.1$, $\mutilde_{\rm p} = 0.8$, and different values of $l/a$. We obtain that the longest period takes place when the prominence thread is centered within the flux tube, i.e., when $L_{\rm e}^-/L = \frac{1}{2} \left( 1 - L_{\rm p}/L \right) = 0.45$ for this particular set of parameters, and $P$ decreases symmetrically around $L_{\rm e}^-/L  =0.45$ when $L_{\rm e}^-/L$ increases or decreases. The dependence of $\td$ on $L_{\rm e}^-/L$ shows the same behavior as $P$. Such as happens with the dependence on $L_{\rm p}/L$ (Fig.~\ref{fig:resapppp}), the dependence on $L_{\rm e}^-/L$ also cancels out when the damping ratio is computed (see Fig.~\ref{fig:displaced1}c). Hence, in our model the value of $\tdp$ is independent of both $L_{\rm p}/L$ and $L_{\rm e}^-/L$. We can also see in Figure~\ref{fig:displaced1}c that the approximate $\tdp$ given by Equation~(\ref{eq:appdampingratiotot2}) remains valid even when the thread is not located at the center of the magnetic tube.

\section{IMPLICATIONS FOR PROMINENCE SEISMOLOGY}
\label{sec:seismology}

\begin{figure}[!tp]
\centering
 \epsscale{0.49}
 \plotone{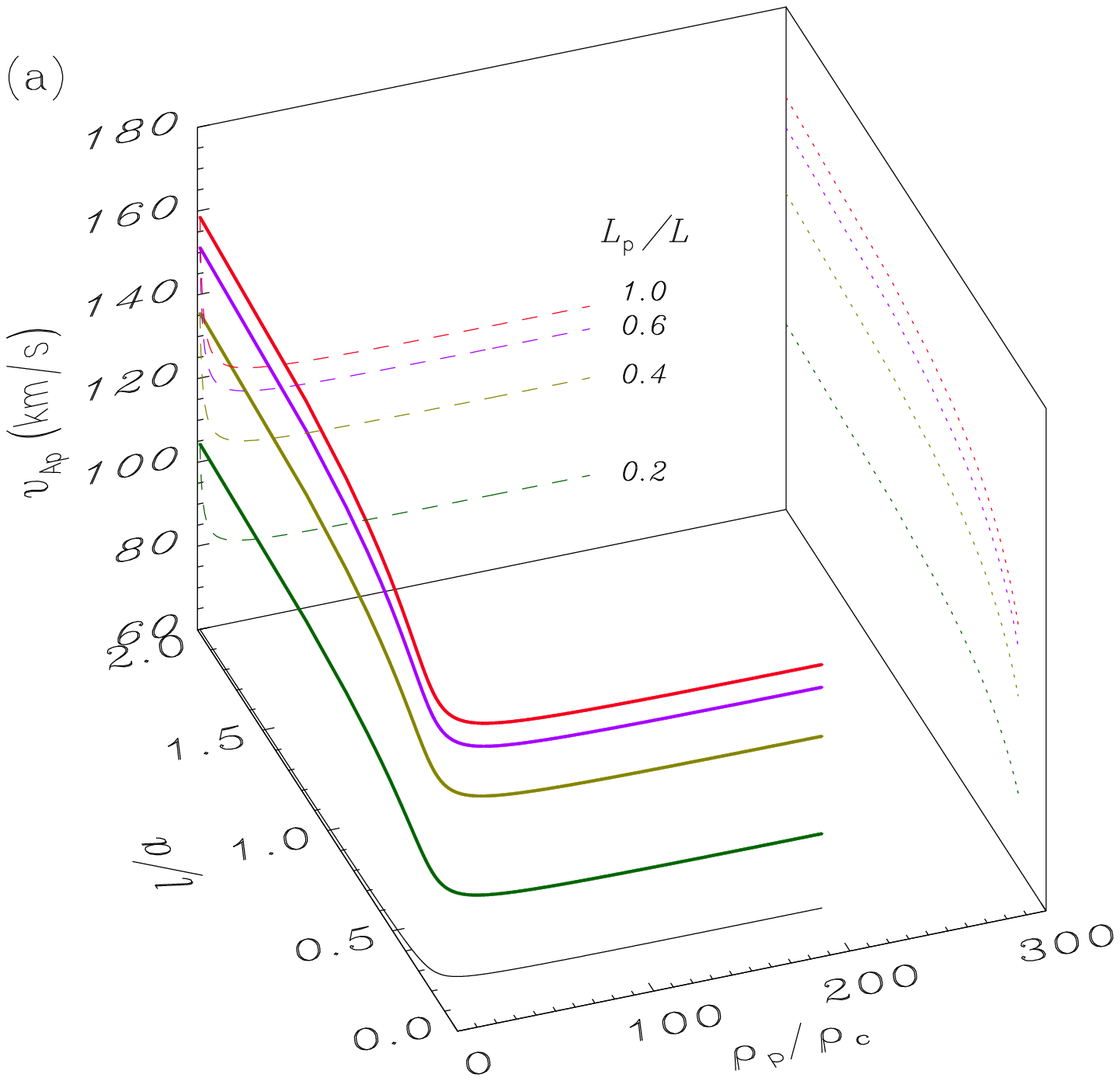}
\plotone{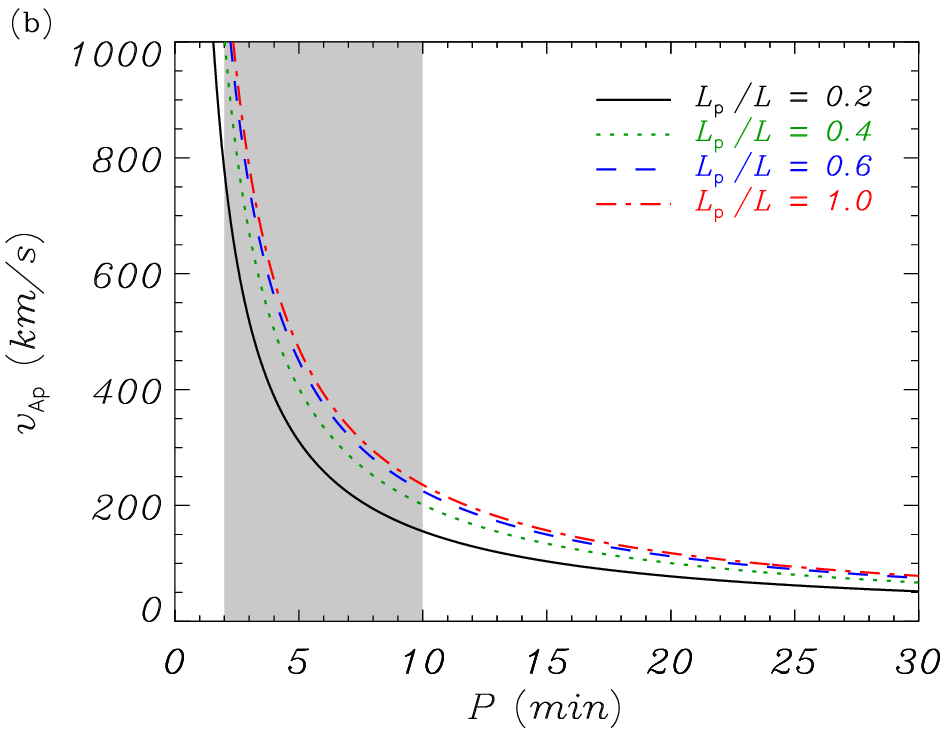}
 \plotone{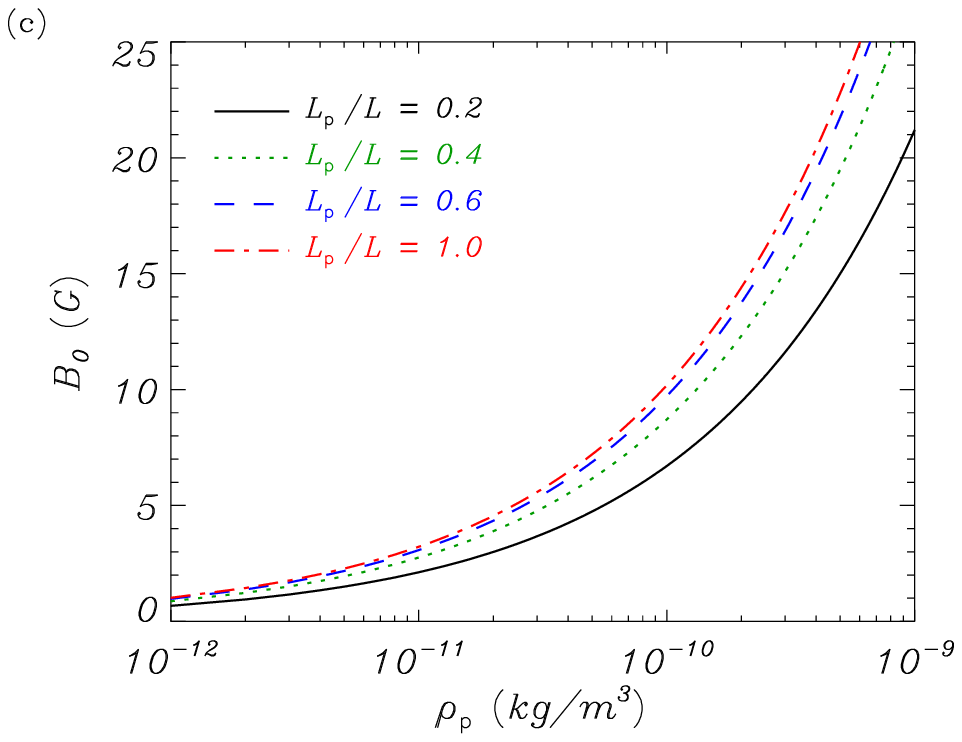}
\caption{(a) Inversion of physical parameters in the ($\rho_{\rm p}/\rho_{\rm c}$, $l/a$, $\vaf$) space for a prominence thread oscillation with $P=20$~min and $\tdp=3$, and for different values of $L_{\rm p}/L$ (indicated within the Figure). The thin continuous, dotted, and dashed lines correspond to the projections of the three-dimensional curves to the ($\rho_{\rm p}/\rho_{\rm c}$, $l/a$)-, ($l/a$, $\vaf$)-, and ($\vaf$, $\rho_{\rm p}/\rho_{\rm c}$)-planes, respectively. (b) Inversion of the prominence Alfv\'en speed, $\vaf$, as a function of the period, $P$, in the limit of large density contrast. The different lines correspond to $L_{\rm p}/L=$~0.2 (solid), 0.4 (dotted), 0.6 (dashed), and 1 (dot-dashed). The shaded zone corresponds to the range of typically observed periods in thread oscillations. (c) Magnetic field strength, $B_0$, as a function of the prominence thread density, $\rho_{\rm p}$, assuming an oscillatory period of $P = 26$~min. The different lines have the same meaning as in panel (b). In all these computations, $L = 10^5$~km.  \label{fig:seis}}
\end{figure}

The obtained results have direct implications for the determination of physical parameters in prominence fine structures using MHD seismology together with observed periods and damping times. The technique of MHD seismology has been previously applied by some authors to obtain information of the plasma physical conditions  in the context of coronal loop oscillations \citep[e.g.,][]{nakaofman,arregui07,arregui08,goossens08}, prominence global oscillations \citep[e.g.,][]{roberts91,regnier,pouget}, and prominence thread oscillations \citep[e.g.,][]{hinode,lin09}.  As partial ionization has a negligible effect on the damping of oscillations we here concentrate on the inversion of parameters using theoretical results for resonantly damped eigenmodes. Following the analytical and numerical inversion schemes by \citet{arregui07} and \citet{goossens08} for coronal loops, \citet{arreguiballester} have recently presented the inversion of prominence Alfv\'en speed, transverse inhomogeneity length-scale, and density contrast in oscillating filament threads using results for resonantly damped kink oscillations in one-dimensional (1D) filament thread models. A similar procedure can be followed with two-dimensional (2D) threads by considering the impact that the length of the thread has on the period and damping ratio of standing kink modes in partially filled fine structure oscillations.

In the following computations, we consider that the prominence thread is located at the center of the magnetic flux tube. In the 2D case and for small $L_{\rm p} / L$, i.e, $L_{\rm p} / L  \lesssim 0.4$, the oscillatory period is approximately given by Equation~(\ref{eq:periodgen2}) for arbitrary density contrast, and by Equation~(\ref{eq:periodgen3}) in the limit of large density contrast. In the 1D case, the kink mode period in the TT approximation is
\begin{equation}
  P \approx \frac{\lambda}{\vaf}  \sqrt{\frac{\rho_{\rm p} + \rho_{\rm c}}{2 \rho_{\rm p}}}, \label{eq:period1d}
\end{equation}
where $\lambda$ is the wavelength. By comparing Equations~(\ref{eq:periodgen2}) and (\ref{eq:period1d}), we see that the expression in the 1D case is equivalent to that in the 2D case if a particular or effective value of $\lambda$ is considered, namely $\lambda \approx \pi  \sqrt{\left( L- L_{\rm p} \right) L_{\rm p}}$. This effective value of $\lambda$ allows us to generalize the 1D inversion to a 2D configuration. Equation~(\ref{eq:periodgen2}) is not accurate enough when $L_{\rm p} / L \gtrsim 0.4$, but we can still use a similar expression for the period for any $L_{\rm p} / L$  as
\begin{equation}
 P \approx C \frac{\pi}{\vaf}  \sqrt{\frac{\rho_{\rm p} + \rho_{\rm c}}{2 \rho_{\rm p}}}  \sqrt{\left( L- L_{\rm p} \right) L_{\rm p}}, \label{eq:period1dcorr}
\end{equation}
with $C$ a correction factor that is computed by comparing the period given by Equation~(\ref{eq:periodgen2}) to the general result of Figure~\ref{fig:ressapp}a (solid line), obtained by solving the full dispersion relation (Equation~(\ref{eq:disperapp})). Thus, $\lambda \approx C \pi  \sqrt{\left( L- L_{\rm p} \right) L_{\rm p}}$ in the general case. For example, by assuming $L=10^5$~km, we obtain $C\approx 1.05$ and $\lambda \approx 9.89 \times 10^4$~km for $L_{\rm p} / L = 0.1$, whereas $C\approx 1.58$ and $\lambda \approx 1.98 \times 10^5$~km for $L_{\rm p} / L = 0.8$. As expected, $\lambda \to 2 L = 2 \times 10^5$~km in the limit $L_{\rm p} / L \to 1$.

  On the other hand, the damping ratio is unaffected by the length of the thread and the same expression for 1D models given by Equation (\ref{eq:appdampingratiotot2}) holds. By following the analytical inversion scheme by \citet{goossens08} the valid equilibrium models that explain equally well a given set of parameters ($P$, $\td$, and $L$) are obtained. The resulting one-dimensional curve in the three-dimensional parameter space is shown in Figure~\ref{fig:seis}a, for different values of the length of the thread. Regardless of the value of $L_{\rm p} / L$, the inversion curve allows us to obtain well constrained values for $\vaf$ and $l/a$ in the limit of high density contrast values. Because of the decrease in $\lambda$ produced by the decrease of the length of the thread, the obtained Alfv\'en speed in the prominence decreases as $L_{\rm p} / L$ gets smaller. An accurate estimate of the length of the thread, in comparison to the length of the magnetic flux tube, is therefore crucial for the determination of the Alfv\'en speed in the thread. As pointed out by \citet{diazperiods}, $L_{\rm p} / L$ can be obtained from the ratio of the fundamental mode period to that of the first harmonic, if these values are reported from the observations. However, because of the independence of the damping ratio on $L_{\rm p} / L$, the projection of the solution curve onto the ($\rho_{\rm p}/\rho_{\rm c}$, $l/a$)-plane remains unaltered regardless of the value of $L_{\rm p} / L$. Hence, the inverted value of the transverse inhomogeneity length-scale, which is obtained taking the limit $\rho_{\rm p}/\rho_{\rm c} \to \infty$, is not affected by different values of the length of the thread. 

The value of $P$ has a direct impact on the seismological determination of the Alfv\'en speed, with the inversion of Figure~\ref{fig:seis}a corresponding to $P=20$~min. The effect of the period on the determination of the prominence Alfv\'en speed is indicated in Figure~\ref{fig:seis}b, where we see that the Alfv\'en speed decreases as the period grows. Here, it is important recalling that the usually reported periods of oscillating threads \citep[e.g.,][]{lin07,lin09,okamoto,ning} are in the range 2 -- 10~min. Note the very large values of $\vaf$ obtained in the range of observed periods (shaded zone in Fig.~\ref{fig:seis}b). By assuming $\rho_{\rm p} = 5\times 10^{-11}$~kg~m$^{-3}$ and $B_0 = 5$~G as typical values of the density and magnetic field strength of quiescent prominences that can be found in the literature, respectively, the corresponding Alfv\'en speed is $\vaf \approx 63$~km/s, which for $L=10^5$~km and $L_{\rm p}/L = 0.2$ gives $P \approx 23.5$~min. Hence, to obtain realistic values of $\vaf$, we have to consider larger periods than those usually observed. Equation~(\ref{eq:periodgen}) indicates that, for the same set of parameters, the period decreases if the thread is displaced from the center of the magnetic tube (see Fig.~\ref{fig:displaced1}a). The maximum shift of the period with respect to the value in the centered case is proportional to $ \sqrt{L_{\rm p}/L}$. Adopting the same parameters as before, the period varies from $P \approx 23.5$~min to $P \approx 9.4$~min when the thread is displaced from the center to the end of the magnetic tube, and so the period enters within the observed range. However, there is no solid basis to assume that all short-period oscillating threads are located at the ends of their magnetic tubes. Alternatively, the presence of flows can also shift the oscillatory period. Although mass flow has not been included in our model, the time-dependent simulations of flowing threads by \citet{hinode} indicate that flow has a minor influence on the period because the flow velocities are much smaller than the Alfv\'en speed. In active region prominences, the larger magnetic field strength could cause larger Alfv\'en speeds, hence thread standing oscillations in active region prominences may have shorter periods than in quiescent prominences. 

Moreover, \citet{lin07,lin09} and \citet{ning} reported that the wavelengths of these short-period oscillations are in the range 700 -- 8,000~km. The  observed wavelengths are between 1 and 2 orders of magnitude smaller than the wavelengths corresponding to standing oscillations computed from Equations~(\ref{eq:period1d}) and (\ref{eq:period1dcorr}). Thus, the reported short wavelengths are impossible to reconcile with the fundamental standing mode of the fine structure. These results suggest that the observed short-period and short-wavelength oscillations of threads in quiescent prominences may not be consistent with an interpretation in terms of standing kink oscillations. A more likely explanation of these short-period and short-wavelength oscillations in terms of propagating kink waves has been performed by \citet{lin09}. These authors seismologically inferred realistic values of the Alfv\'en speed and magnetic field strength by assuming a propagating wave interpretation \citep[see details in][]{lin09}. 

The limited duration of the currently available Doppler time series may prevent the observation of standing modes with periods larger than 10~min \citep[e.g., the time series last for only 18~min in the observations of][]{lin07}. However, there are a few evidences of larger periods in longer time series that could be consistent with standing modes.  These longer periods have been obtained from Doppler signals that have been averaged over a large area. As the spatial scales of standing modes are very large, their oscillatory patterns could be found in spatially averaged signals. On the contrary, the averaging process could mix signals coming from different adjacent threads, and so there is no full confidence that the period corresponds to an individual thread oscillation. \citet{yi} detected thread oscillations with period of 16~min and minimum wavelength of $ 2 \times 10^4$~km in their Doppler observations with low spatial resolution ($1''$), while \citet{lin04} reported 26~min and minimum wavelength of $4 \times 10^4$~km in their averaged Doppler signals. Although not only the periods but also the wavelengths reported in these two works are consistent with a standing oscillation, we have to be very cautious with this interpretation. Additional information as, e.g., the polarisation of the oscillations and the phase difference between perturbations, would be needed for a more robust analysis and the unequivocal determination of the wave mode.

To perform a simple application, let us assume that the period of 26~min reported by \citet{lin04} corresponds to a standing thread oscillation. Then, the estimation of the magnetic field strength in the prominence thread is possible. Following the analysis presented by \citet{lin09} using the observational period and the seismologically determined Alfv\'en speed (as in Fig.¬\ref{fig:seis}a), the magnetic field strength can be computed for a given prominence density. This result is plotted in Figure~\ref{fig:seis}c. For a typical density of $\rho_{\rm p}=5\times 10^{-11}$~kg~m$^{-3}$, magnetic field strengths in the range 4~--~7 G are obtained, approximately, when varying $L_{\rm p} / L$ between 0.2 and 1. These values of the magnetic field strength are in agreement with previous magnetic field measurements in quiescent prominences using the Hanle effect \citep[e.g.,][]{leroy}. Hence, the method can be applied in the future using real data if reliable observations of standing thread oscillations are reported.

\section{CONCLUSION}

\label{sec:conclusion}

In this paper, we have investigated standing kink oscillations of prominence fine structures. The longitudinal nonuniformity has been taken into account by modeling the fine structure as a magnetic tube only partially filled with the prominence material. We have followed an analytical method based on the TT approximation and have found a dispersion relation for kink oscillations damped by Cowling's diffusion and resonant absorption in the TB approach. This dispersion relation has been numerically solved and a parametric study of the solution has been performed. In addition to the general dispersion relation, we have obtained simple analytical approximations to the period, the damping time,  and their ratio. 

Both approximate and full results conclude that resonant absorption is much more efficient than Cowling's diffusion for the kink mode damping, with the values of $\td / P$ in agreement with those reported in the observations. As happens for long-wavelength propagating waves in the dense part of the fine structure \citep{solerRAPI}, the prominence plasma ionization degree turns out to be irrelevant for the resonant damping of the oscillations. In addition, the value of $\td / P$ is found to be independent of both the position of the prominence thread within the magnetic tube and the length of the prominence region, and coincides with the value for a homogeneous and fully ionized prominence tube \citep{arregui08,solerslow}. 

Finally, we have discussed the seismological implications of our analytical results, in particular Equations~(\ref{eq:periodgen3}) and (\ref{eq:tdsimp}). With these expressions, it is possible to estimate some relevant physical parameters of oscillating threads if the values of the period and damping time are available from the observations. Following this idea, we have performed a seismological inversion of the prominence thread Alfv\'en speed and the transverse inhomogeneity length-scale by using our theoretical results and adopting ad-hoc values for the period and damping time. We have shown that for short-period (2 -- 10~min) and short-wavelength (700 -- 8,000~km) thread oscillations, the determined Alfv\'en speeds are much larger than the expected, realistic values, pointing out that short-period and short-wavelength thread oscillations may not be consistent with a standing kink mode interpretation and could be related to propagating waves. On the contrary, thread oscillations with periods larger than 10~min and wavelengths larger than $10^4$~km may be interpreted as standing oscillations. In this last case, the Alfv\'en speed and magnetic field strength estimated by the seismological inversion are realistic in the context of prominences. Thus, the method can be put into practice to extract indirect information of prominences when standing thread oscillations are unequivocally observed and the oscillation parameters, i.e., period, damping time, and wavelength, along with the thread length are provided from the observations.

In this work, we have assumed that there is an abrupt jump of the density at the boundary between the prominence thread and the evacuated part of the magnetic tube. This simplification has allowed us to proceed analytically. Actually, one should expect a continuous variation of the plasma properties in the longitudinal direction between both regions, which could affect somehow our present results. The study of the damping of kink oscillations in fully nonuniform two-dimensional fine structures is broached numerically by \citet{arregui2d} in a following investigation. Other additional ingredients as, e.g., magnetic twist or curvature and the presence of flows might be included in future works.

\acknowledgements{
We acknowledge the unknown referee for valuable comments. RS thanks Marcel Goossens for some useful discussions and for the kind hospitality during his stay in Leuven, where part of this work was performed. RS also thanks Jaume Terradas for helpful suggestions. The authors acknowledge the financial support received from the Spanish MICINN and FEDER funds (AYA2006-07637). The authors also acknowledge discussion within ISSI Team on Solar Prominence Formation and Equilibrium: New data, new models. RS thanks the CAIB for a fellowship.}

\appendix

\section{DERIVATION OF THE DISPERSION RELATION}
\label{appendix}

Here, we give extensive details about the method that leads us to the dispersion relation (Equation~(\ref{eq:disperappcomp})).

\subsection{Boundary conditions at $r=a$}

First, we must consider appropriate boundary conditions for the solutions of  Equations~(\ref{eq:ptin}) and (\ref{eq:pout}) at the cylinder edge, i.e., $r=a$. In the evacuated part of the tube, we assume $\rho_{\rm e} = \rho_{\rm c}$ and there is no transverse transitional layer. Hence, the boundary conditions are those given by \citet{dymovaruderman} in their Equation~(4), namely
\begin{equation}
 \left[ \left[ p_{\rm T} \right] \right]= 0, \quad \left[ \left[ v_r \right]\right] = 0, \quad \textrm{at} \quad r= a \quad \textrm{for} \quad |z| > L_{\rm p}/2, \label{eq:jumpeva}
\end{equation}
where $\left[ \left[ X \right]\right]$ stands for the jump of the quantity $X$. 

On the other hand, in the prominence part of the tube we consider the effect of resonant absorption in the transitional layer. We follow the treatment by \citet{andries2005}, who generalize the concept of the jump conditions at the resonance position of \citet{SGH91} to the case of a longitudinally inhomogeneous tube. \citet{andries2005}  combined the jump conditions with the TB approximation, i.e, $l/a \ll 1$, to obtain analytical expressions of the dispersion relation and the frequency for longitudinally stratified tubes. The accuracy of this analytical method was numerically verified by \citet{arregui05}, who found a good agreement between the expressions of \citet{andries2005} and their numerical computations.  In the thin tube approximation, i.e., $a/L \ll 1$,  \citet{dymovaruderman2} follow a similar formalism and also provide equivalent expressions for the jump conditions that can be applied to our perturbations. Hence, it is convenient to express $p_{\rm T}$ and $v_r$ as
\begin{equation}
 p_{\rm T} = \sum_{n=1}^\infty p_{{\rm T}n} G_n, \qquad v_r = \sum_{n=1}^\infty v_{rn} G_n, \label{eq:expansion}
\end{equation}
where $p_{{\rm T}n}$ and $v_{rn}$ are the coefficients of the series expansions of $p_{\rm T}$ and $v_r$, respectively, with respect to the functions $G_n$ determined by the Sturm-Liouville problem
\begin{equation}
 \va^2 \left(r\right) \frac{{\rm d}^2 G_n}{{\rm d} z^2} =  -  \lambda_n^2\left(r\right)  G_n,  \label{eq:gtrans}
\end{equation}
with appropriate boundary conditions for $G_n$ at $z = \pm L_{\rm p}/2$, with $\lambda_n^2$ the corresponding eigenvalues. Equation~(\ref{eq:gtrans}) describe the spectrum of Alfv\'en modes, with $\lambda_n\left(r\right)$ the corresponding frequencies of the Alfv\'en continuum. In general, it is not straight-forward to deduce the boundary conditions for $G_n$ at $z = \pm L_{\rm p}/2$ because they are given by the continuity of $G_n$ at $z = \pm L_{\rm p}/2$, and the value of $G_n$ at $z = \pm L_{\rm p}/2$ is also determined by the properties of the evacuated region. In a longitudinally homogeneous tube, i.e., for $L_{\rm p} = L$ and with the Alfv\'en speed depending on the radial direction only, we simply have that the boundary conditions are $G_n \left( \pm L/2 \right) = 0$ and obtain $\lambda_n\left(r\right)  \equiv \omega_{\rm A}\left(r \right) = \frac{n \pi}{L} \va \left(r \right)$, with $n=1$, 2, \dots

In our notation, the jump conditions for $p_{{\rm T}n}$ and $v_{rn}$ provided by  \citet{dymovaruderman2} are
\begin{equation}
  \left[ \left[ p_{{\rm T}n} \right] \right]= 0, \quad \left[ \left[ v_{rn} \right]\right] = - \pi \omega_{\rm R}  \frac{m^2 / a^2}{ \left| \rho_0 \Delta_n \right|_{r_{{\rm A}n}}} p_{{\rm T}n} , \quad \textrm{at} \quad r= r_{{\rm A}n} \quad \textrm{for} \quad |z| < L_{\rm p}/2, \label{eq:jumpdense}
\end{equation}
where $r_{{\rm A}n}$ is the Alfv\'en resonance position for the $n$th mode and $\Delta_n = \frac{{\rm d}}{{\rm d} r} \left( \omega_{\rm R}^2 - \lambda_n^2  \right)$, with $\omega_{\rm R}$ the real part of the frequency. According to Equation~(\ref{eq:expansion}), the condition for $p_{{\rm T}n}$ in Equation~(\ref{eq:jumpdense}) leads to $\left[ \left[ p_{\rm T} \right] \right]= 0$ at $r= r_{{\rm A}n}$ for $|z| < L_{\rm p}/2$. On the contrary, the condition for $v_{rn}$ depends on $\left| \rho_0 \Delta_n \right|_{r_{{\rm A}n}}$, and a more detailed analysis is needed.

For our subsequent analysis, we do not need the precise value of $ \lambda_n\left(r\right)$ but only its functional dependence on the radial direction. For the given sinusoidal density profile in the transitional layer, we can express the Alfv\'en speed squared in the transitional layer as $\va^2 \left(r\right) = \vaf^2 / f \left(r\right) $, with
\begin{equation}
  f \left(r\right) = \frac{1}{2} \left\{\left(1+\frac{\rho_{\rm c}}{\rho_{\rm p}}\right) - \left( 1-\frac{\rho_{\rm c}}{\rho_{\rm p}}\right)\sin \left[\frac{\pi}{l}\left( r-a\right)\right]\right\}.
\end{equation}
Hence, Equation~(\ref{eq:gtrans}) is rewritten as
\begin{equation}
 \vaf^2 \frac{{\rm d}^2 G_n}{{\rm d} z^2} =  -  \lambda_n^2\left(r\right) f \left(r\right) G_n.  \label{eq:gtrans2}
\end{equation}
With no loss of generality, we can assume that $G_n$ is only a function of $z$, i.e., the different magnetic surfaces are not coupled to each other. So, according to Equation~(\ref{eq:gtrans2}), the quantity $\lambda_n^2\left(r\right) f \left(r\right)$ corresponds to the Alfv\'en eigenvalue squared in the prominence part of the tube. Since $\rho_{\rm p}$ is homogeneous, its corresponding Alfv\'en eigenvalue does not depend on $r$, meaning that the radial contribution of $\lambda_n^2\left(r\right)$ and $f \left(r\right)$ cancel out. Thus, we define $\lambda_{{\rm p} n}^2 \equiv \lambda_n^2\left(r\right) f \left(r\right)$, with $\lambda_{{\rm p} n}$ a constant corresponding to the Alfv\'en eigenvalue in the homogeneous prominence thread. Therefore, we have
\begin{equation}
  \lambda_n^2\left(r\right) =   \frac{\lambda_{{\rm p} n}^2}{f \left(r\right)}, \label{eq:lambdan}
\end{equation}
where all the radial dependence of $\lambda_n^2\left(r\right)$ comes from the function $f \left(r\right)$. With the help of Equation~(\ref{eq:lambdan}), we obtain that $ \Delta_n = \lambda_n^2\left(r\right) f' \left(r\right) / f \left(r\right)$, where the prime denotes the radial derivative. Finally, we use the resonant condition, namely $ \lambda_n^2\left( r_{{\rm A}n} \right) = \omega_{\rm R}^2$, and write
\begin{equation}
  \left| \rho_0 \Delta_n \right|_{r_{{\rm A}n}} = \omega_{\rm R}^2   \left| \pd_r \rho_0 \right|_{r_{{\rm A}n}}. \label{eq:rescond}
\end{equation}
 Thus, the condition for $v_{rn}$ in Equation~(\ref{eq:jumpdense}) becomes
\begin{equation}
  \left[ \left[ v_{rn} \right]\right] =  - \pi \frac{m^2 / a^2}{\omega_{\rm R}   \left| \pd_r \rho_0 \right|_{r_{{\rm A}n}} } p_{{\rm T}n} , \quad \textrm{at} \quad r= r_{{\rm A}n} \quad \textrm{for} \quad |z| < L_{\rm p}/2. \label{eq:jumpdense2}
\end{equation}
The value of $r_{{\rm A}n}$ is in principle different for each value of $n$ and could be determined from the resonant condition $ \lambda_n^2\left( r_{{\rm A}n} \right) = \omega_{\rm R}^2$ if the eigenvalues $ \lambda_n^2\left( r \right) $ were {\em a priori} known and the number of Alfv\'en eigenmodes that are resonant to the kink mode is also known. Hence, the derivative of the density profile at each resonant position could be computed. A reasonable assumption in the TT and TB limits is to consider $r_{{\rm A}n} \approx a$ for all $n$, so that $ \left| \pd_r \rho_0 \right|_a$ is a constant independent of $n$, meaning that we are assuming that all resonances take place at the same position. For our sinusoidal profile, $ \left| \pd_r \rho_0 \right|_a \approx \pi \left( \rho_{\rm p} - \rho_{\rm c}   \right) / 2 l$.  Note that this is not a very strong restriction for the fundamental kink mode, since it is likely that the resonant condition is satisfied for $n=1$ only, because $\lambda_n^2\left( r \right)$ grows as $n$ increases and so only one resonance takes place. The approximation $r_{{\rm A}n} \approx a$ might not be valid for the kink mode overtones, but here we restict ourselves to the fundamental mode. Alternatively, a simpler linear density profile could be adopted \citep{goossens02} in which the derivative of the density profile does not depend on the position. Therefore and using Equation~(\ref{eq:expansion}), we arrive at 
\begin{equation}
  \left[ \left[ v_{r} \right]\right] = - \pi \frac{m^2 / a^2}{\omega_{\rm R}   \left| \pd_r \rho_0 \right|_a } p_{{\rm T}} , \quad \textrm{at} \quad r= a \quad \textrm{for} \quad |z| < L_{\rm p}/2. \label{eq:jumpdense3}
\end{equation}
For $L_{\rm p} = L$ the jump condition of Equation~(\ref{eq:jumpdense3}) consistingly reduces to that provided by \citet{SGH91}.

\subsection{Solution in the evacuated regions}

Let us consider first the boundary conditions for the evacuated parts (Equation~(\ref{eq:jumpeva})). The analysis here is identical to that of \citet{dymovaruderman}. For the condition on the total pressure perturbation, we obtain $A_{\rm e} \left(z\right) = A_{\rm c} \left(z\right)=A \left(z\right)$. Next, we rewrite Equation~(\ref{eq:ptvr}) as
\begin{equation}
 \vazero^2 \frac{\pd^2 v_r}{\pd z^2} + \omega^2 v_r = - \frac{i \omega}{\rho_0} \frac{\pd p_{\rm T}}{\pd r}. \label{eq:vrpteva}
\end{equation}
We evaluate Equation~(\ref{eq:vrpteva}) for $r \approx a$ on both sides of the tube boundary. Thus, in the evacuated part,
\begin{equation}
 \vae^2  \frac{\pd^2 v_{r \rm e}}{\pd z^2} + \omega^2 v_{r \rm e} = - \frac{i \omega}{\rho_{\rm e}} \frac{m}{a} A \left(z\right), \quad \textrm{for} \quad r \lessapprox a, \label{eq:vre}
\end{equation}
whereas in the corona,
\begin{equation}
 \vac^2  \frac{\pd^2 v_{r \rm c}}{\pd z^2} + \omega^2 v_{r \rm c} =  \frac{i \omega}{\rho_{\rm c}} \frac{m}{a} A \left(z\right), \quad \textrm{for} \quad r \gtrapprox a. \label{eq:vrc}
\end{equation}
According to the boundary condition for $v_r$ given by Equation~(\ref{eq:jumpeva}), $v_{r \rm e} = v_{r \rm c}$. Thus, we combine Equations~(\ref{eq:vre}) and (\ref{eq:vrc}) to find the following two expressions
\begin{equation}
 \frac{\pd^2 v_r}{\pd z^2} = i \omega \frac{m}{a} \frac{\mu}{B_0^2}  \frac{\rho_{\rm e} + \rho_{\rm c}}{\rho_{\rm e} - \rho_{\rm c}}   A \left(z\right), \label{eq:d2vrfin} \\
\end{equation}
\begin{equation}
 v_r = - i \frac{m}{a} \frac{1}{\omega} \frac{2}{ \rho_{\rm e} - \rho_{\rm c}} A \left(z\right), \label{eq:vrfin}
\end{equation}
where we have considered that the magnetic field is homogeneous. Now, we differentiate Equation~(\ref{eq:vrfin}) with respect to $z$ twice and compare the resulting expression with 
Equation~(\ref{eq:d2vrfin}). We obtain
\begin{equation}
 \frac{{\rm d}^2 A \left( z \right)}{{\rm d} z^2} + \frac{\omega^2}{c_{k \rm e}^2} A \left(z\right) = 0. \label{eq:azeva}
\end{equation}
The quantity $c_{k \rm e}$ corresponds to the kink speed in the evacuated region (Equation~(\ref{eq:kinkspeed})). Note that in our particular application $ \rho_{\rm e} = \rho_{\rm c}$, so $c_{k \rm e} = \vae$. However, we keep the general notation $c_{k \rm e}$ in the following expressions.

To solve Equation~(\ref{eq:azeva}), we consider the line-tying condition at the photosphere, i.e., $A \left(z^\pm_{\rm wall} \right) = 0$. Therefore, the solution in the two evacuated zones is
\begin{equation}
 A \left(z\right) = \left\{
\begin{array}{lll}
 C_1 \sin \left[  \frac{\omega}{c_{k \rm e}} \left( z - L_{\rm p}/2 - L_{\rm e}^+ \right)  \right], & \textrm{for} & z > L_{\rm p}/2, \\ \\
C_2 \sin \left[  \frac{\omega}{c_{k \rm e}} \left( z + L_{\rm p}/2 + L_{\rm e}^- \right)  \right], & \textrm{for} & z < -L_{\rm p}/2,
\end{array}
\right.  \label{eq:soleva}
\end{equation}
where $C_1$ and $C_2$ are constants.

\subsection{Solution in the prominence thread}

In the prominence thread, we adopt the TB approach and use the jump conditions given by Equation~(\ref{eq:jumpdense}) as our boundary conditions. The combination of both formalisms to investigate resonant waves in coronal flux tubes has been reviewed by \citet{goossensrev}. Again, the condition over the total pressure perturbation gives $A_{\rm p} \left(z\right) = A_{\rm c} \left(z\right)=A \left(z\right)$. Near the boundary we express $v_{r \rm c} = v_{r \rm p} + \delta v_r$, where $\delta v_r$ is the jump of the radial velocity perturbation provided by Equation~(\ref{eq:jumpdense3}), namely
\begin{equation}
 \delta v_r =  - \pi  \frac{m^2 / a^2}{\omega_{\rm R}   \left| \pd_r \rho_0 \right|_a } p_{\rm T}.
\end{equation}
As before, we evaluate Equation~(\ref{eq:ptvr}) on both sides of the tube boundary and, after some algebra, we arrive at the following expressions
\begin{eqnarray}
  \frac{\pd^2 v_{r \rm p}}{\pd z^2} &=&  \frac{m}{a} \frac{i \omega}{ \vac^2 -  \Gamma_{\rm A p}^2 }  \frac{\rho_{\rm p} + \rho_{\rm c}}{\rho_{\rm p} \rho_{\rm c}}  A \left(z\right) \nonumber \\ &+&  \frac{m^2/a^2}{ \omega_{\rm R}   \left| \pd_r \rho_0 \right|_a} \frac{\pi}{  \vac^2 -  \Gamma_{\rm A p}^2 } \left[ \vac^2 \frac{{\rm d}^2 A \left( z \right)}{{\rm d} z^2} + \omega^2 A \left(z\right)   \right], \label{eq:d2vrfin2} 
\end{eqnarray}
\begin{eqnarray}
 v_{r \rm p} &=& -  \frac{m}{a} \frac{i}{\omega} \frac{  \rho_{\rm p} \Gamma_{\rm A p}^2 + \rho_{\rm c} \vac^2 }{\rho_{\rm p} \rho_{\rm c} \left(\vac^2 -  \Gamma_{\rm A p}^2  \right)} A \left(z\right) \nonumber \\ &-& \frac{\pi}{\omega^2}  \frac{\Gamma_{\rm A p}^2 }{\vac^2 - \Gamma_{\rm A p}^2 }  \frac{m^2 / a^2}{ \omega_{\rm R}   \left| \pd_r \rho_0 \right|_a} \left[ \vac^2 \frac{{\rm d}^2 A \left( z \right)}{{\rm d} z^2} + \omega^2 A \left(z\right)   \right]. \label{eq:vrfin2}
\end{eqnarray}
Now, we differentiate Equation~(\ref{eq:vrfin2}) with respect to $z$ twice and compare the resulting expression with Equation~(\ref{eq:d2vrfin2}), obtaining
\begin{eqnarray}
  \frac{{\rm d}^4 A \left( z \right)}{{\rm d} z^4} &+& \left[ \omega^2 \left( \frac{\Gamma_{\rm A p}^2 + \vac^2}{\Gamma_{\rm A p}^2 \vac^2} \right) + \frac{i \omega}{\pi} \frac{ \omega_{\rm R}   \left| \pd_r \rho_0 \right|_a}{m/a} \left(  \frac{\rho_{\rm p} \Gamma_{\rm A p}^2 + \rho_{\rm c} \vac^2 }{\rho_{\rm p} \rho_{\rm c} \Gamma_{\rm A p}^2 \vac^2} \right) \right] \frac{{\rm d}^2 A \left( z \right)}{{\rm d} z^2} \nonumber \\
&+& \omega^2 \left[ \frac{\omega^2 }{\Gamma_{\rm A p}^2 \vac^2} + \frac{i \omega}{\pi} \frac{ \omega_{\rm R}   \left| \pd_r \rho_0 \right|_a}{m/a} \left(  \frac{\rho_{\rm p} + \rho_{\rm c} }{\rho_{\rm p} \rho_{\rm c} \Gamma_{\rm A p}^2 \vac^2} \right) \right]  A \left( z \right) = 0. \label{eq:atot}
\end{eqnarray}
 The general Equation~(65) of \citet{dymovaruderman2} and our Equation~(\ref{eq:atot}) are equivalent if a constant piecewise density is assumed in the former and Cowling's diffusion is omitted in the latter. Equation~(\ref{eq:atot}) can be solved by taking a solution of the form $\exp \left( i k_z z \right)$ and obtaining the subsequent forth-order polynomial for $k_z$. Two independent values of $k_z$ are possible, namely $k_{z1}$ and $k_{z2}$. Thus, the general solution of Equation~(\ref{eq:atot}) is
\begin{equation}
 A \left( z \right) = D_1 \exp \left( i k_{z1} z \right) + D_2 \exp \left(- i k_{z1} z \right) +  D_3 \exp \left( i k_{z2} z \right) + D_4 \exp \left(- i k_{z2} z \right),
\end{equation}
with $D_1$, $D_2$, $D_3$, and $D_4$ constants that are determined by the boundary conditions at $z = \pm L_{\rm p}/2$. However, to keep this general analysis implies that the following expressions are complicated and require an additional mathematical effort. Instead, we choose a more restrictive way to simplify matters. 

For our next analysis, Equation~(\ref{eq:atot}) is rewritten in a convenient form as
\begin{eqnarray}
 \frac{b^2}{\omega^2} \frac{{\rm d}^4 A \left( z \right)}{{\rm d} z^4} + \frac{{\rm d}^2 A \left( z \right)}{{\rm d} z^2} + \frac{\omega^2}{\tilde{c}_{k \rm p}^2}  A \left( z \right) = 0, \label{eq:simp}
\end{eqnarray}
where $\tilde{c}_{k \rm p}^2$ and $b^2$ are defined in Equations~(\ref{eq:ckfull1}) and (\ref{eq:bfull}). 

In the case of the fundamental mode, one could assume that, when the terms related to Cowling's diffusion and resonant absorption are present, the characteristic scale for the variations of the eigenfunctions in the $z$-direction is only slightly modified with respect to the ideal case without transitional layer. Therefore, a reasonable approximation is to relate the forth-order derivative of $A \left(z\right)$ in Equation~(\ref{eq:simp}) with the second-order derivative as follows
\begin{equation}
  \frac{{\rm d}^4 A \left( z \right)}{{\rm d} z^4} \sim - \mathcal{K}^2  \frac{{\rm d}^2 A \left( z \right)}{{\rm d} z^2},
\end{equation}
where the quantity $\mathcal{K}$ plays the role of the longitudinal wavenumber. We can approximate $\mathcal{K}$ by its expression in the ideal case, namely
\begin{equation}
 \mathcal{K}^2 \approx \frac{\omega^2}{c_{k \rm p}^2},
\end{equation}
with $c_{k \rm p}^2$ the ideal kink speed. Hence, Equation~(\ref{eq:simp}) becomes
\begin{eqnarray}
  \frac{{\rm d}^2 A \left( z \right)}{{\rm d} z^2} +  \frac{\omega^2 }{\tilde{c}_{k \rm p}^2 \left( 1 - \frac{b^2}{c_{k \rm p}^2} \right)}  A \left( z \right) \approx 0, \label{eq:azdens}
\end{eqnarray}
which is formally identical to Equation~(\ref{eq:azeva}). It is important recalling that the approximation of the forth-order $z$-derivative of $A \left(z\right)$  may introduce some uncertainty in the solutions of Equation~(\ref{eq:azdens}) in comparison with the solutions of the full Equation~(\ref{eq:atot}). However, we expect a minor discrepancy in the case of the fundamental mode because its characteristic scale in the $z$-direction should not be essentially modified when the terms related to Cowling's diffusion and resonant absorption are taken into account in the equations.

The solution of Equation~(\ref{eq:azdens}) is
\begin{equation}
 A \left(z\right) = E_1 \cos \left( \frac{\omega}{\tilde{c}_{k \rm p}\sqrt{ 1 - \frac{b^2}{c_{k \rm p}^2} }} z \right) + E_2 \sin \left( \frac{\omega}{\tilde{c}_{k \rm p}\sqrt{ 1 - \frac{b^2}{c_{k \rm p}^2} }} z \right),  \qquad \textrm{if} \qquad |z| \leq L_{\rm p}/2, \label{eq:soldense}
\end{equation}
with $E_1$ and $E_2$ constants. When the prominence thread is centered within the magnetic tube, i.e., $L_{\rm e}^- = L_{\rm e}^+ = \frac{1}{2} \left( L - L_{\rm p} \right)$, the solutions of Equation~(\ref{eq:azdens}) can be separated according to their symmetry about $z = 0$. Thus, even modes are described by $E_1 \ne 0$, $E_2 = 0$, and odd modes by $E_1 = 0$, $E_2 \ne 0$.

\subsection{Matching the solutions at $z=\pm L_{\rm p} / 2$}

In order to match the solution in the prominence thread (Equation~(\ref{eq:soldense})) with those in the evacuated regions (Equation~(\ref{eq:soleva})), we impose the boundary conditions
\begin{equation}
 \left[ \left[ A \right]\right] = 0, \qquad \left[ \left[  \frac{{\rm d} A }{{\rm d} z} \right]\right] = 0, \qquad \textrm{at} \qquad z = \pm L_{\rm p}/2,
\end{equation}
corresponding to a contact discontinuity \citep{goed}. After applying these boundary conditions, the general dispersion relation (Equation~(\ref{eq:disperappcomp})) is finally obtained.

\end{document}